\documentclass{article}
\usepackage[utf8]{inputenc}
\usepackage[margin=1in]{geometry}
\usepackage{subcaption}
\usepackage{graphicx}
\usepackage{amsmath}
\usepackage[hyphens]{url}
\usepackage{hyperref}
\hypersetup{colorlinks=false,breaklinks=true,hidelinks}
\usepackage{cleveref}
\usepackage{apacite}
\usepackage{natbib}
\usepackage{authblk}
\usepackage{multirow}
\usepackage{booktabs}
\usepackage{makecell}
\graphicspath{{figures/}}
\usepackage{lineno}
\usepackage{enumitem}
\setlist{nosep}

\title{Robust Earthquake Location using Random Sample Consensus (RANSAC)}
\author[1]{Weiqiang Zhu\thanks{zhuwq@berkeley.edu}}
\author[1]{Bo Rong}
\author[2]{Yaqi Jie}
\author[2]{S. Shawn Wei}
\affil[1]{Department of Earth and Planetary Science, University of California, Berkeley, Berkeley, CA, USA}
\affil[2]{Department of Earth and Environmental Sciences, Michigan State University, MI, USA}

\date{}

\begin{document}

\maketitle

\begin{abstract}
Accurate earthquake location, which determines the origin time and location of seismic events using phase arrival times or waveforms, is fundamental to earthquake monitoring. While recent deep learning advances have significantly improved earthquake detection and phase picking, particularly for smaller-magnitude events, the increased detection rate introduces new challenges for robust location determination. These smaller events often contain fewer P- and S-phase picks, making location accuracy more vulnerable to false or inaccurate picks.
To enhance location robustness against outlier picks, we propose a machine learning method that incorporates the Random Sample Consensus (RANSAC) algorithm. RANSAC employs iterative sampling to achieve robust parameter optimization in the presence of substantial outliers. By integrating RANSAC's iterative sampling into traditional earthquake location workflows, we effectively mitigate biases from false picks and improve the robustness of the location process. We evaluated our approach using both synthetic data and real data from the Ridgecrest earthquake sequence. The results demonstrate comparable accuracy to traditional location algorithms while showing enhanced robustness to outlier picks.
\end{abstract}

\section{Introduction}

Earthquake source parameters, including time, location, and mechanism, provide foundational information for understanding earthquake characteristics, fault zone structures, and various subsurface processes associated with volcanoes, glaciers, geothermal energy, oil and gas extraction, and CO$_2$ sequestration \citep{kanamori2004physics,ellsworth2013injection,faulkner2010review,shelly2016fluid}. To construct earthquake catalogs and deliver early warning information, seismic networks routinely detect earthquakes and pick phase arrivals to determine origin times and hypocenters \citep{kanamori2005real,allen2009status,hutton2010earthquake}.

The basic mechanism of earthquake location is to find source parameters that best fit observed phase arrival times across a network of stations, based on a velocity model and a travel-time calculation algorithm. The classic approach, i.e., Geiger's method \citep{geiger1912probability}, employs iterative least-squares optimization by linearizing travel-time changes relative to small perturbations in earthquake hypocenters and applies Gauss-Newton nonlinear optimization \citep{thurber1985nonlinear}. Commonly used earthquake location programs such as HYPO71 \citep{lee1975hypo71} and HYPOINVERSE \citep{klein2002users} build upon Geiger's method and operate routinely in earthquake monitoring systems. These programs incorporate various improvements to enhance location quality, including sophisticated weighting schemes for phase arrival qualities, optimized hyperparameters for iteration and damping, and location error estimations.

Advanced algorithms have been developed to accommodate inaccurate velocity models and improve optimization methods with uncertainty quantification. Important techniques include directly inverting earthquake location and velocity model \citep{lahr1979hypoellipse,thurber1983earthquake,kissling1994initial,thurber1992hypocentervelocity}, adding static station corrections \citep{douglas1967joint,ellsworth1975bear,frohlich1979efficient,pujol1988comments,frohlich1979efficient,lomax2005reanalysis} and source-specific station terms \citep{richards2000earthquake,shearer2005southern,lin2006comploca,lin2018source,lomax2022high}, using equal differential travel-times at station pairs \citep{zhou1994rapid,font2004hypocentre}, using double-difference travel-times for earthquake pairs \citep{waldhauser2000doubledifference,waldhauser2001hypodd,trugman2017growclust}, and including location uncertainty estimation using Bayesian optimization \citep{hirata1987maximumlikelihood,myers2007bayesian,lomax2009earthquake,smith2022hyposvi}. These advancements have significantly improved earthquake location accuracy, enabling detailed seismicity mapping with precise spatiotemporal information for studying complex earthquake sequences, fault zone structures, induced seismicity, volcanic earthquakes, and other geological processes \citep{waldhauser2002fault,waldhauser2008large,hauksson2012waveform,richards2000earthquake,shelly2020high,ross20203d,park2022basement,wilding2023magmatic,trugman2024high}.

In addition to these improvements in location accuracy, a critical challenge lies in enhancing location robustness against outliers in arrival-time measurements. This issue has become increasingly significant as detection thresholds continue to lower through advanced detection algorithms such as deep learning-based phase picking models, identifying more earthquake signals from continuous seismic archives \citep{ross2018p,mousavi2020earthquake,zhu2019phasenet,sun2023phase}. A fundamental trade-off exists between increasing detection numbers and minimizing false positives throughout the earthquake monitoring workflow, from phase detection/picking, phase association, to event location.
In the phase detection/picking step, efforts to detect weak signals from small earthquakes that are often near or below noise levels inevitably generate false picks from anthropogenic, instrumental, and natural noise sources. Additionally, arrival time tends to exhibit increased errors in low signal-to-noise ratio (SNR) waveforms, making precise first-arrival picking increasingly challenging.
In the phase association step, which clusters picks across seismic networks into individual earthquakes while filtering out false picks, two challenges arise from increasing false picks: First, small earthquakes typically yield limited P- and S-phase arrivals, necessitating lower thresholds for minimum pick numbers per event, which, in turn, increases the chance that false picks may be associated to create false event detections. Second, abundant false picks from background noise increase the likelihood of false picks coincidentally matching true event travel-time moveouts, potentially biasing location results, especially for events with few picks.
These errors from phase picking and association steps, including false picks and inaccurate phase arrival times, can accumulate to the earthquake location step, potentially resulting in erroneous or biased location results.

Several approaches have been proposed to improve location robustness, including alternative target functions less sensitive to outliers (e.g., $L1$-norm loss or Huber loss instead of $L2$-norm loss) \citep{anderson1982robust,shearer1997improving,trugman2017growclust}, likelihood functions incorporating uncertainties \citep{lomax2000probabilistic,smith2022hyposvi}, and specialized residual weighting and outlier filtering schemes in iterative optimization \citep{klein2002users}.
In this work, we propose combining the Random Sample Consensus (RANSAC) algorithm \citep{fischler1981random} with earthquake location to enhance robustness against outliers. 
The RANSAC algorithm is designed for robust parameter estimation in the presence of significant outliers.
Instead of minimizing error across all data points, RANSAC aims to maximize the inlier consensus set through randomly sampling multiple subsets that are sufficient to constrain model parameters, thereby effectively isolating and disregarding outliers. 
RANSAC has been widely applied in computer vision tasks \citep{hartley2003multiple,szeliski2022computer} and has recently emerged in geophysical applications, including recognizing linear fault planes \citep{fadakar2013ransac,pollitz2017geodetic,liu2018enhanced,skoumal2019characterizing} and solving phase picking and association tasks \citep{zhu2016automatic,woollam2020hex,zhu2017estimation,zhu2021multichannel}. 
These studies directly apply the RANSAC algorithm based on linear curve or hyperbolic curve fitting \citep{zhu2017estimation}.
\citep{trugman2020spatiotemporal} demonstrated improved laboratory earthquake location accuracy by combining grid-searching with RANSAC for outlier pick removal.

Building on RANSAC's advantages, we present an earthquake location method that integrates RANSAC's random sampling iteration with traditional optimization processes to achieve robust outlier resistance. We validate our method using both synthetic examples and field data from the 2016 Ridgecrest sequence. The improved robustness to false picks in the location step would enable lower thresholds in earlier steps of phase picking and association for small-magnitude earthquake detection. The enhanced accuracy in earthquake locations ultimately contributes to higher-quality earthquake catalogs, supporting a wide range of geophysical studies.

\section{Methods}

\subsection{Earthquake location}

Earthquake localization aims to determine source parameters, origin time and hypocenter, that best fit observed phase arrival times across a network of stations, assuming a velocity model and travel-time calculation function.
Travel-times of phase arrivals $t$ are commonly estimated using ray-tracing \citep{um1987fast,aki1976determination,zhao1992tomographic,vcerveny2020seismic} or by solving the eikonal equation \citep{vidale1988finite,van1991upwind,podvin1991finite,tong2021adjoint,smith2020eikonet}.
In this work, we solve the eikonal equation to calculate travel-time:
\begin{align}
\left|\nabla t(x, y, z)\right| &= \frac{1}{v(x, y, z)} \\
t(x_0, y_0, z_0) &= 0
\end{align}
where $t(x, y, z)$ is the travel-time from hypocenter to station, $v(x, y, z)$ is the velocity model, $t_0$ is the earthquake origin time, and $x_0, y_0, z_0$ are the earthquake hypocenter coordinates.
We employ the fast-sweeping method to solve the eikonal equation \citep{zhao2005fast}, which offers both computational efficiency and straightforward implementation.
The phase arrival-time is expressed as the origin time plus travel-time, with an optional station correction term $\delta_s$:
\begin{align}
\hat{T} = t_0 + t(x, y, z) + \delta_s
\end{align}
The objective for earthquake location is to minimize the misfit between observed ($T_i$) and calculated ($\hat{T}_i$) phase arrival-times:
\begin{align}
\mathcal{L} = \sum_{i=1}^{N} w_i f \left(\hat{T}_i - T_i \right)
\label{eq:loss}
\end{align}
where $N$ is the number of phase measurements, $w_i$ is the weight assigned to each phase pick, and $f$ is the loss function. Common loss functions include $L2$-norm loss, $L1$-norm loss, and huber loss \citep{huber1964robust}.
We select the Huber loss function in this study, as it is less sensitive to outliers, same as the $L1$-norm loss and converges quickly for small residuals, akin to the $L2$-norm loss:
\begin{align}
f(x) = 
\begin{cases}
\frac{1}{2} x^2 & \text{if } |x| \leq \delta \\
\delta \cdot (|x| - \frac{1}{2} \delta) & \text{if } |x| > \delta
\end{cases}
\end{align}
The earthquake location process involves minimizing the target function $\mathcal{L}$ to optimize the earthquake origin time $t_0$ and hypocenter ($x_0, y_0, z_0$):
\begin{align}
\min_{t_0, x_0, y_0, z_0} \mathcal{L} = \sum_{i=1}^{N} w_i f \left(t_0 + t(x, y, z) + \delta_s - T_i\right)
\label{eq:target}
\end{align}
Given the non-linearity of the travel-time function, we employ the BFGS quasi-Newton optimizer \citep{nocedal1999numerical,byrd1995limited} to optimize the target function.
The gradient of the target function with respect to hypocenter coordinates can be computed either by interpolating the spatial derivatives of the travel-time field ($\partial t / \partial x$, $\partial t / \partial y$, $\partial t / \partial z$) or through automatic differentiation techniques \citep{paszke2017automatic,baydin2018automatic,zhu2021general}. The gradient with respect to origin time is linear and computed analytically.

\subsection{RANSAC-based optimization}

\begin{figure}
    \centering
    \includegraphics[width=0.6\linewidth]{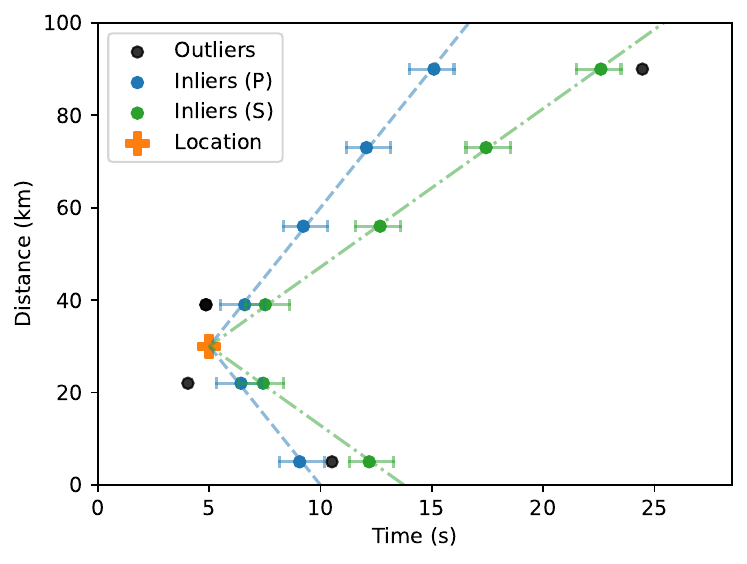}
    \caption{Schematic diagram of the RANSAC-based earthquake location algorithm, which employs iterative sampling to identify and exclude outlier picks while optimizing earthquake location parameters.}
    \label{fig:randac}
\end{figure}

The earthquake location process described above assumes reliable picks with residual errors following a Gaussian distribution (for $L2$-norm loss) or a Laplace distribution (for $L1$-norm loss). However, phase picks are often contaminated by outliers, especially for small-magnitude earthquakes or in noisy conditions.
To improve robustness against outlier picks, we employ the Random Sample Consensus (RANSAC) algorithm (\Cref{fig:randac}).
The RANSAC algorithm can be summarized in the following steps:
\begin{enumerate}
    \item Randomly sample a subset of phase picks $S$ from the group of picks $G$;
    \item Evaluate subset quality (e.g., number of P and S picks); if insufficient, resample a new subset;
    \item Locate earthquake using the subset $S$ by minimizing target function \Cref{eq:target}; \label{step:inversion}
    \item Apply the model to the group of picks $G$ and compute evaluation metrics (e.g., the number of inliers and residuals); \label{step:evaluate}
    \item Evaluate model quality; Update the best model if current model is superior;
    \item Repeat for $K$ iterations or until meeting predefined stopping criteria (e.g., maximum iterations, maximum number of inliers, or minimum residuals);
    \item Use the best inlier subset to determine final earthquake location.
\end{enumerate}

RANSAC's iterative sampling relies on the assumption that a small subset of picks can adequately constrain model parameters. This assumption is valid for earthquake location because only four free parameters, i.e., origin time and hypocenter coordinates, need to be determined, while the number of picks per event typically exceeds four.
Physical constraints in earthquake location inform the selection of RANSAC hyperparameters, such as minimum subset size and maximum allowable residuals. For example, we set the minimum subset size as $\max(S_{min}, \alpha G)$, where $S_{min}$ is the minimum number of picks, and $\alpha$ represents the approximate ratio of true picks within $G$; and we set the maximum allowable time residuals to 1 second. These hyperparameters can be optimized based on specific application scenarios.

\subsection{Other improvements}

Since RANSAC's iterative sampling process operate independently of the earthquake location optimization, enhancements to earthquake location methods can seamlessly integrate with RANSAC. We demonstrate two improvements: static station corrections and phase amplitude-based outlier detection.

\subsubsection{Station correction terms}

Static station correction terms are used to account for systematic errors in observed phase arrival-times related to station-specific velocity model inaccuracies. These correction terms can be estimated from residuals between observed and calculated arrival-times for each station:
\begin{align}
    \Delta \delta_s = \sum_{i=1}^{N} w_i \left(t_0 + t(x, y, z) + \delta_s - T_i\right) / \sum_{i=1}^{N} w_i
\end{align}
where $N$ is the number of phase picks, $w_i$ is the pick weight, and $\Delta \delta_s$ is the correction term for station $s$. We estimate separate station correction terms for P and S phases at each station.

\subsubsection{Phase amplitudes for outlier detection}

Phase amplitudes, while less precise for constraining locations, can identify outliers in phase picks. Amplitude information enhances phase association \citep{zhu2022earthquake} by distinguishing picks with similar arrival-times but differing amplitudes from earthquakes of varying magnitudes and distances.
In conditions with numerous false picks, outliers near travel-time moveouts may bias earthquake locations, but their inconsistent amplitudes can help identify and remove them.
During Step \ref{step:evaluate} of RANSAC, we calculate both travel-time and amplitude residuals:
\begin{align}
    \log A = c_0 + c_1 M + c_2 \log R
\end{align}
where $A$ is phase amplitude, $M$ is earthquake magnitude, $R$ is epicentral distance, and $c_0$, $c_1$, and $c_2$ are coefficients. Integrating this linear relationship into RANSAC enables outlier detection with minimal computational overhead.

\section{Results}

To benchmark the performance of the RANSAC-based earthquake location method, we evaluated it using both synthetic examples and the 2016 Ridgecrest earthquake sequence.

\subsection{Benchmarking on synthetic datasets}

\begin{figure}
    \centering
    \includegraphics[width=0.6\linewidth]{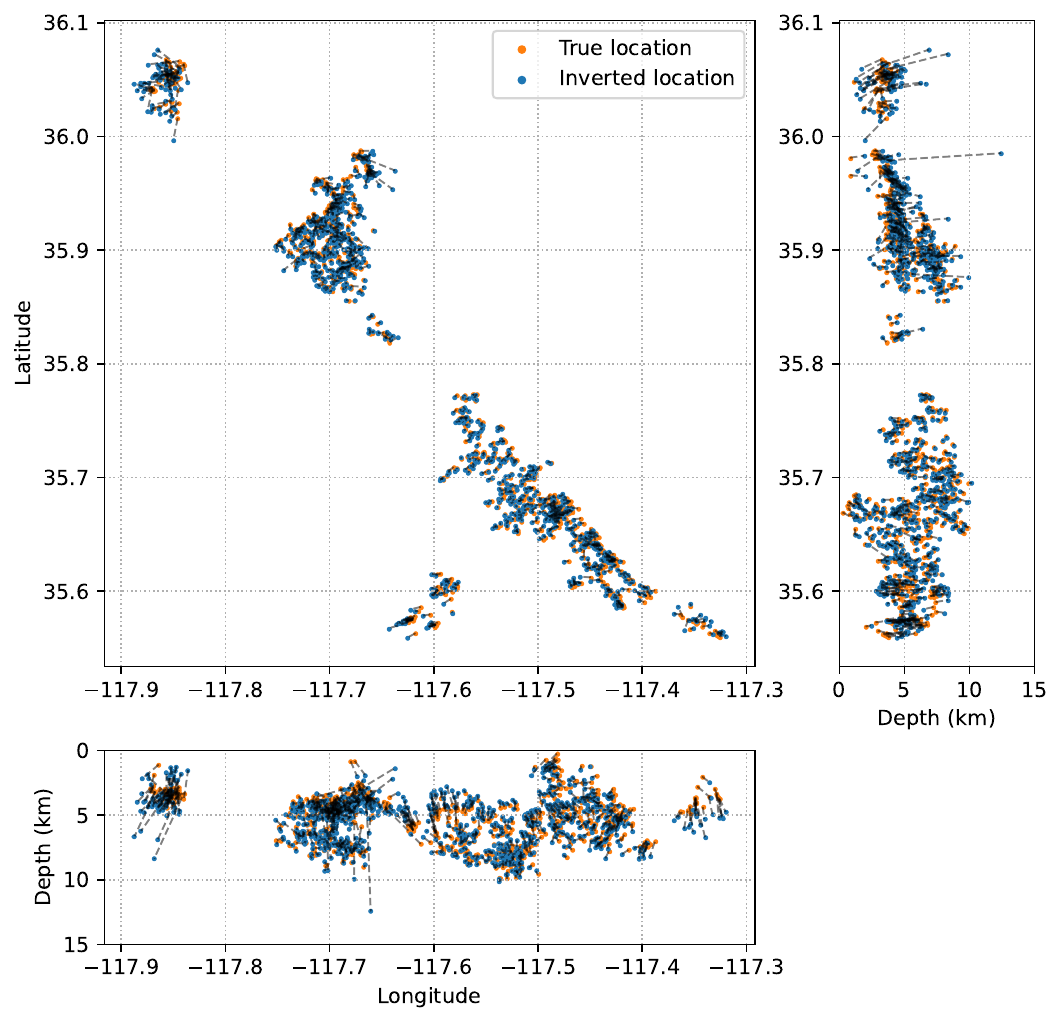}
    \caption{Comparison of estimated and true earthquake locations in the synthetic dataset.}
    \label{fig:synthetic_test}
\end{figure}

\begin{figure}
    \centering
    \includegraphics[width=0.7\linewidth]{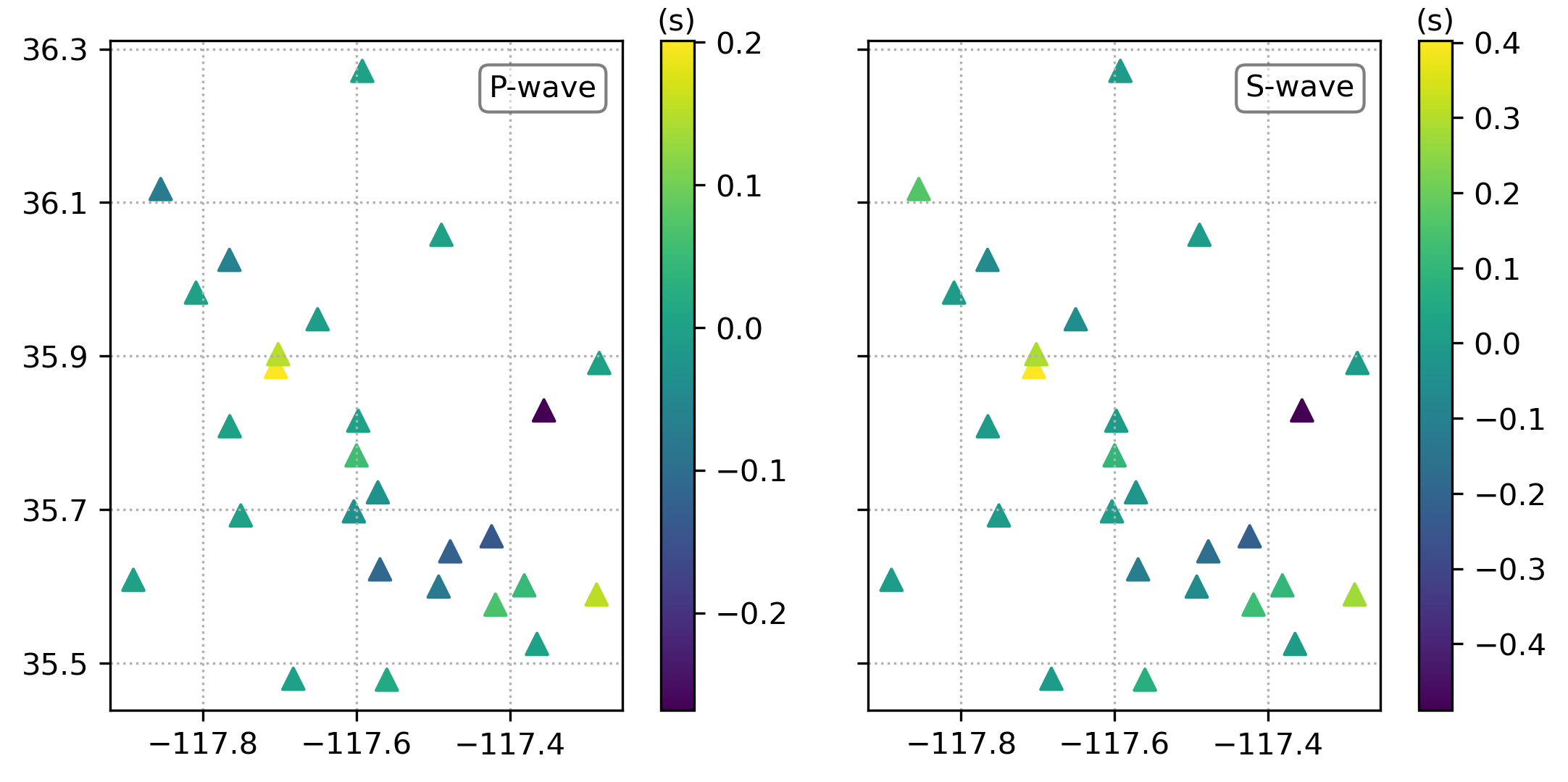}
    \caption{Estimated static station terms for P-wave and S wave.}
    \label{fig:sst}
\end{figure}

\begin{figure}
    \centering
    \includegraphics[width=0.7\linewidth]{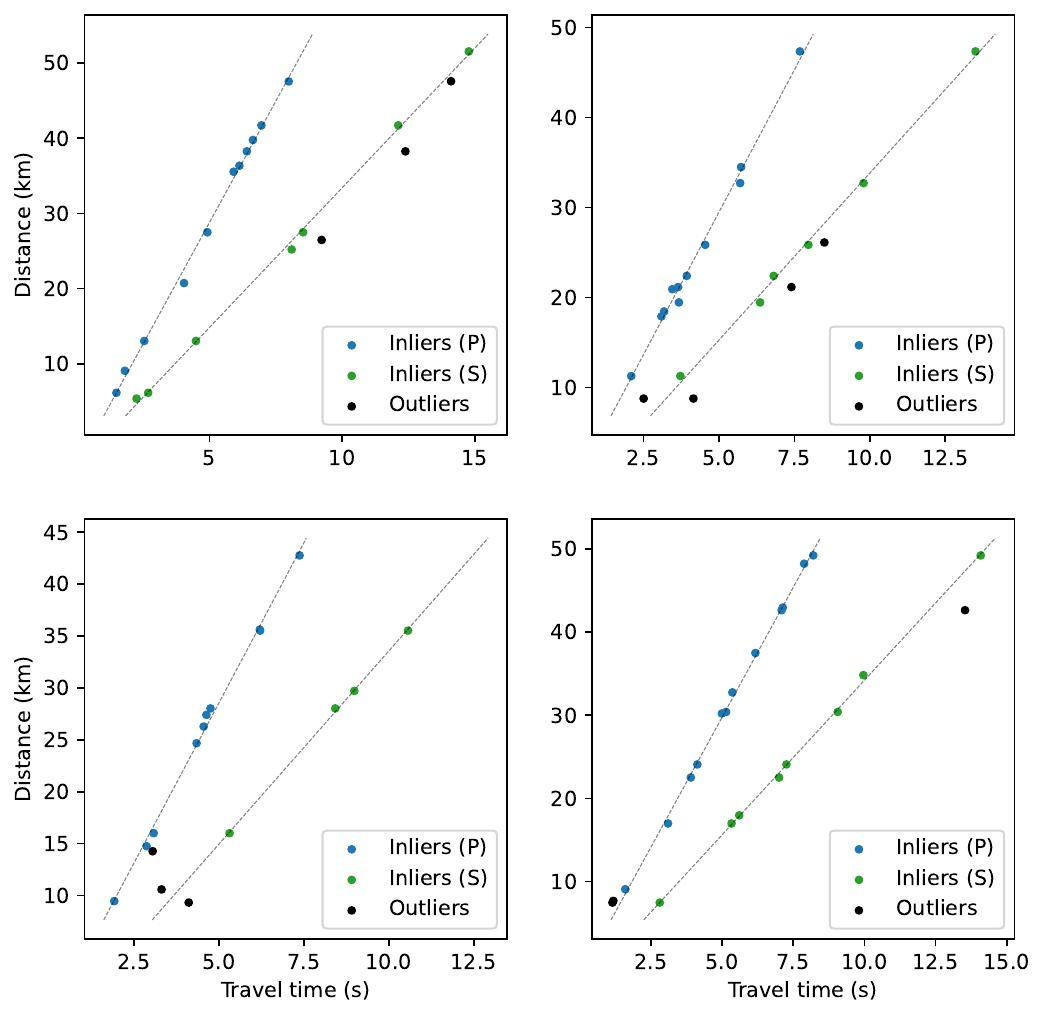}
    \caption{Examples of identified outlier picks.}
    \label{fig:outlier}
\end{figure}

We first evaluated the earthquake location performance using a controlled synthetic experiment developed by \citet{yu2024accuracy}.
The synthetic datasets of phase picks were generated using the fast-marching method \citep{white2020pykonal} with a 3D velocity model, incorporating realistic picking errors, phase availabilities, and phase outliers to simulate real-world conditions.
In their controlled experiment, \citet{yu2024accuracy} compared the performance of commonly used earthquake location algorithms, including Hypoinverse, VELEST, NonLinLoc, NonLinLoc\_SSST, and HypoSVI.
This synthetic dataset provides an ideal benchmark for evaluating the RANSAC-based earthquake location method, which uses the fast-sweeping method \citep{zhao2005fast} for travel-times calculations and the BFGS optimization method for earthquake location invertion.
\Cref{fig:synthetic_test,fig:sst,fig:outlier} show the ground-truth earthquake locations alongside the inverted locations; the estimated static station terms (SST), and filtered outliers using RANSAC, respectively.
\Cref{tab:synthetic_test} presents a quantitative evaluation of the effects of adding static station terms and RANSAC iterations, compared with other commonly used location methods.

The synthetic dataset results validate the effectiveness of both static station terms and RANSAC iterations.
First, the vanilla ADLoc results demonstrate that our implementation using the fast-sweeping eikonal solver for phase travel-time calculations and the BFGS quasi-Newton optimizer for earthquake location optimization works correctly, though as expected, achieves lower accuracy than well-optimized algorithms like Hypoinverse, VELEST, and NonLinLoc.
Second, the ADLoc+SST results show that static station terms effectively reduce location bias from inaccurate velocity models, improving accuracy to levels comparable to or better than other location algorithms. The estimated SST for P-phase and S-phase are shown in \Cref{fig:sst}.
Third, the ADLoc+SST+RANSAC results achieve the best performance through additional outlier pick removal. Step \ref{step:evaluate} of the RANSAC iterations effectively detects and removes outlier picks based on customizable thresholds.
Without phase amplitude information in the benchmark dataset, we relied only on travel-time residuals for outlier detection. The threshold for maximum travel-time residual, a hyperparameter, can be adjusted based on pick quality. Experiments with thresholds of 1.0s and 0.2s both improved location results, with the tighter 0.2s threshold achieving better performance on this relatively clean benchmark dataset.
According to \citet{yu2024accuracy}, the dataset contains 1\% P-phase and 4\% S-phase outliers among the total 19,994 phase picks.
Using a 0.2s threshold, ADLoc detected 184 (0.9\%) P-phase and 539 (2.7\%) S-phase outliers, matching the known outlier proportions. Examples of detected outliers are shown in \Cref{fig:outlier}.

\begin{table}[]
\centering
\caption{Quantitative comparison of location accuracy between different methods on the synthetic dataset. Lower values indicate better performance.}
\label{tab:synthetic_test}
\resizebox{0.8\textwidth}{!}{%
\begin{tabular}{lccc}
\hline
Method                  & \multicolumn{2}{c}{Mean Accuracy Error (km)} & Chamfer Distance \\ \cline{2-3}
                        & Horizontal              & Depth              &                  \\ \hline
HypoInverse             & 0.824                   & 1.118              & 1.617            \\
VELEST                  & 0.696                   & 0.559              & 1.170            \\
NonLinLoc               & 0.953                   & 0.969              & 1.626            \\
ADLoc                   & 1.132                   & 1.257              & 1.802            \\
ADLoc+SST               & 0.680                   & 0.539              & 1.080            \\
ADLoc+SST+RANSAC (1.0s) & 0.630                   & 0.498              & 1.035            \\
ADLoc+SST+RANSAC (0.2s) & 0.456                   & 0.450              & 0.915            \\ \hline
\end{tabular}%
}
\end{table}

\subsection{Applications to real data}

\begin{figure}
    \centering
    \begin{subfigure}{0.8\linewidth}
        \hfill
        \includegraphics[width=\linewidth]{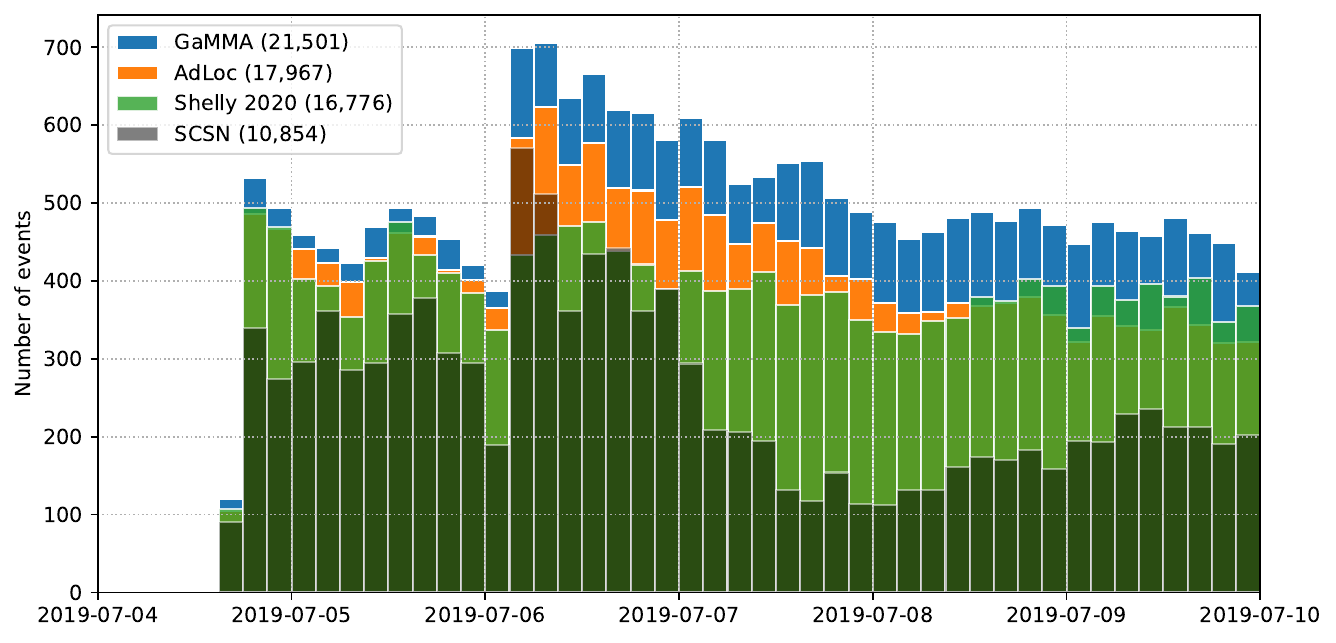}
        \caption{}
    \end{subfigure}
    \begin{subfigure}{0.8\linewidth}
        \hfill
        \includegraphics[width=0.985\linewidth]{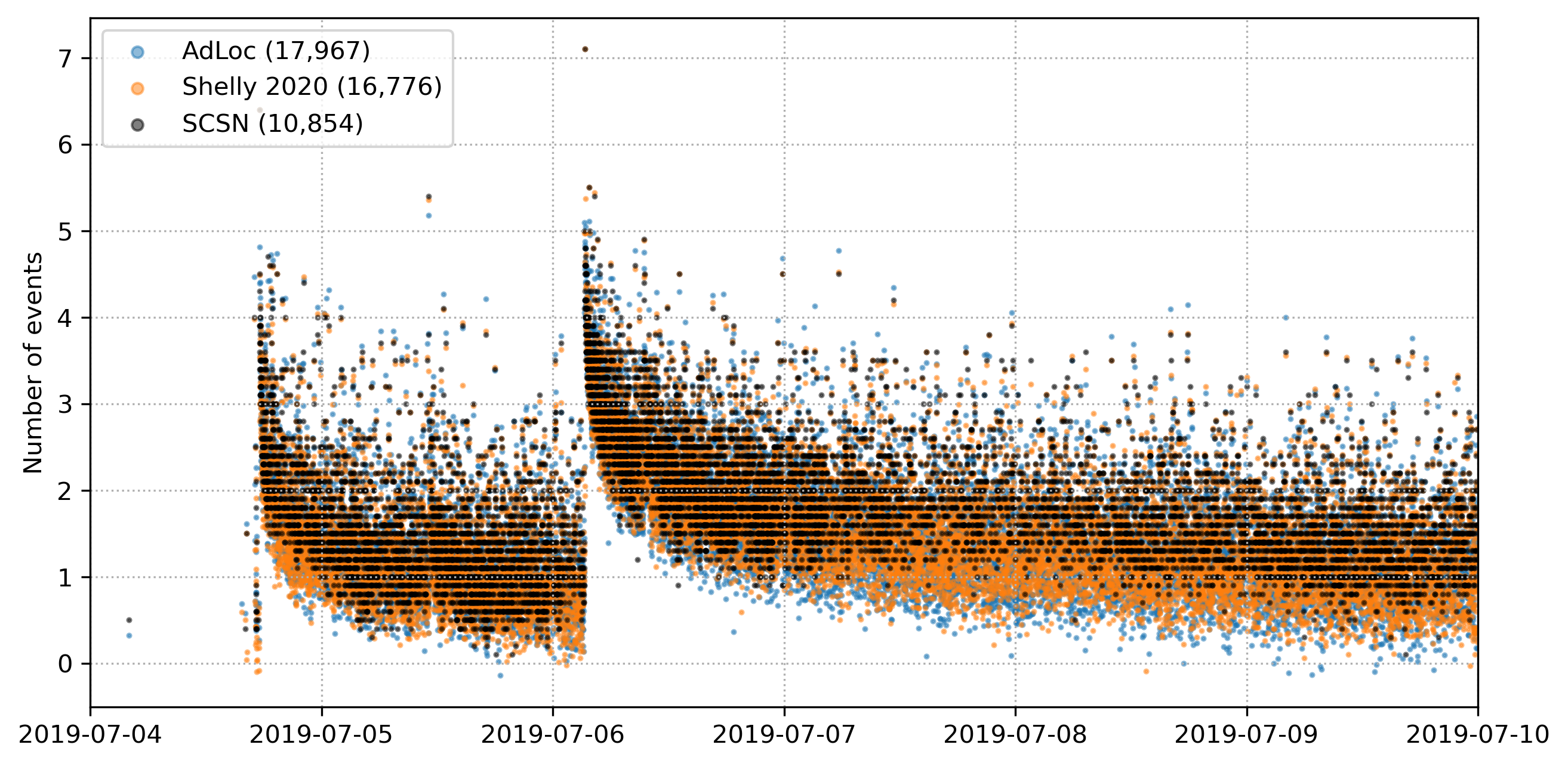}
        \caption{}
    \end{subfigure}
    \caption{Temporal distribution of detected earthquake events during the Ridgecrest sequence: (a) temporal distribution of earthquake frequency; (b) magnitude-time distribution of detected events.}
    \label{fig:event_time}
    \end{figure}

\begin{figure}
        \centering
        \includegraphics[width=0.5\linewidth]{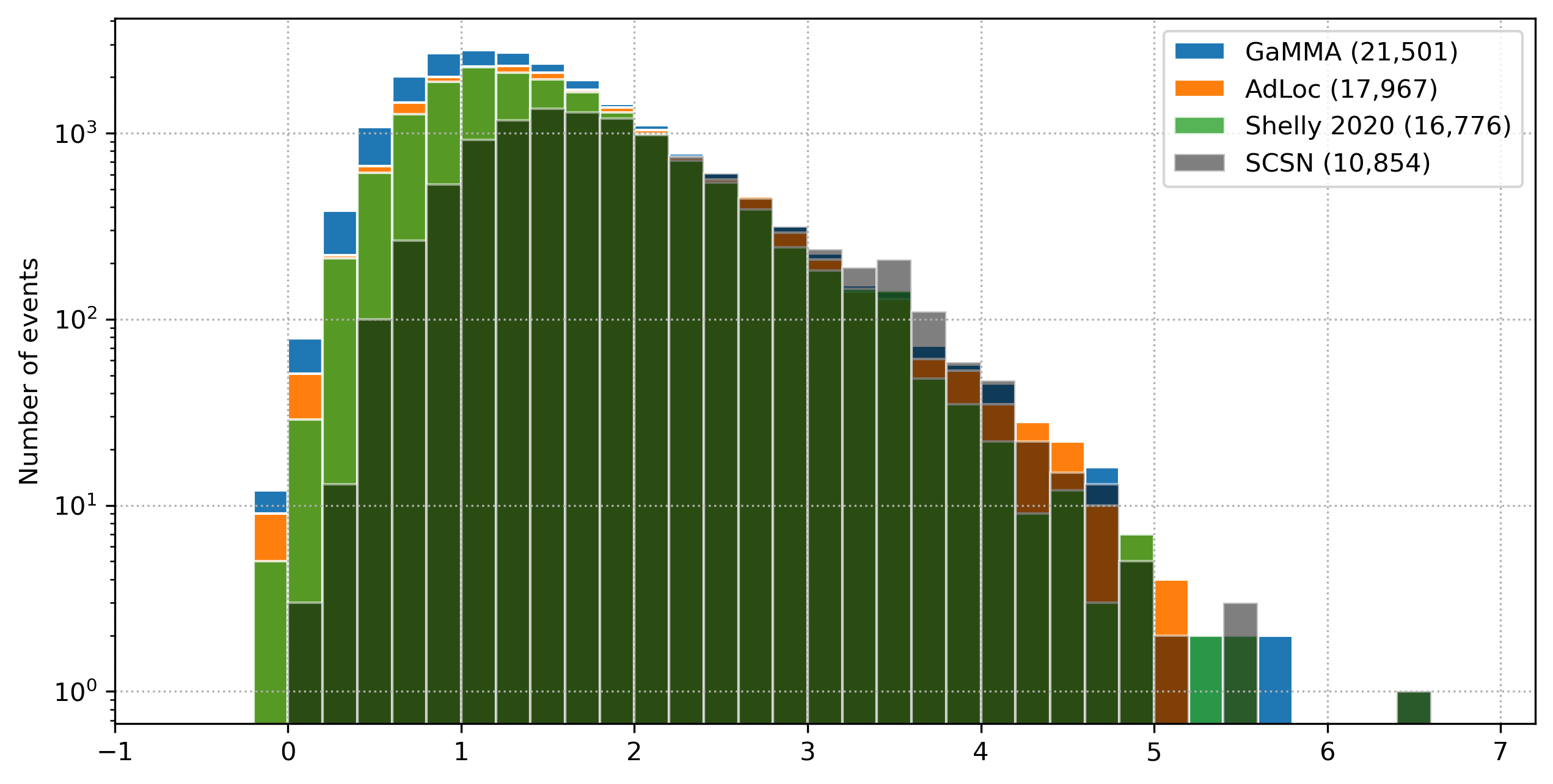}
        \caption{Magnitude distribution of detected events during the Ridgecrest sequence.}
        \label{fig:event_mag}
    \end{figure}

\begin{figure}
        \centering
        \includegraphics[width=\linewidth]{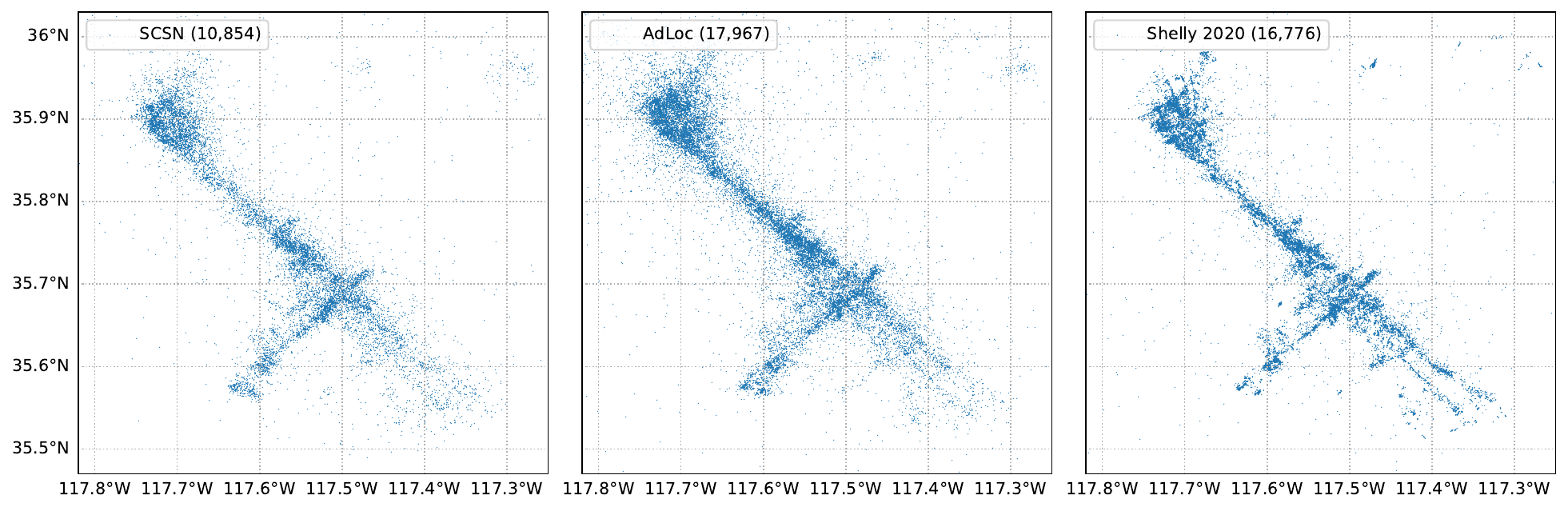}
        \caption{Spatial distribution of located events during the Ridgecrest sequence..}
        \label{fig:event_location}
    \end{figure}

\begin{figure}
        \centering
        \includegraphics[width=\linewidth]{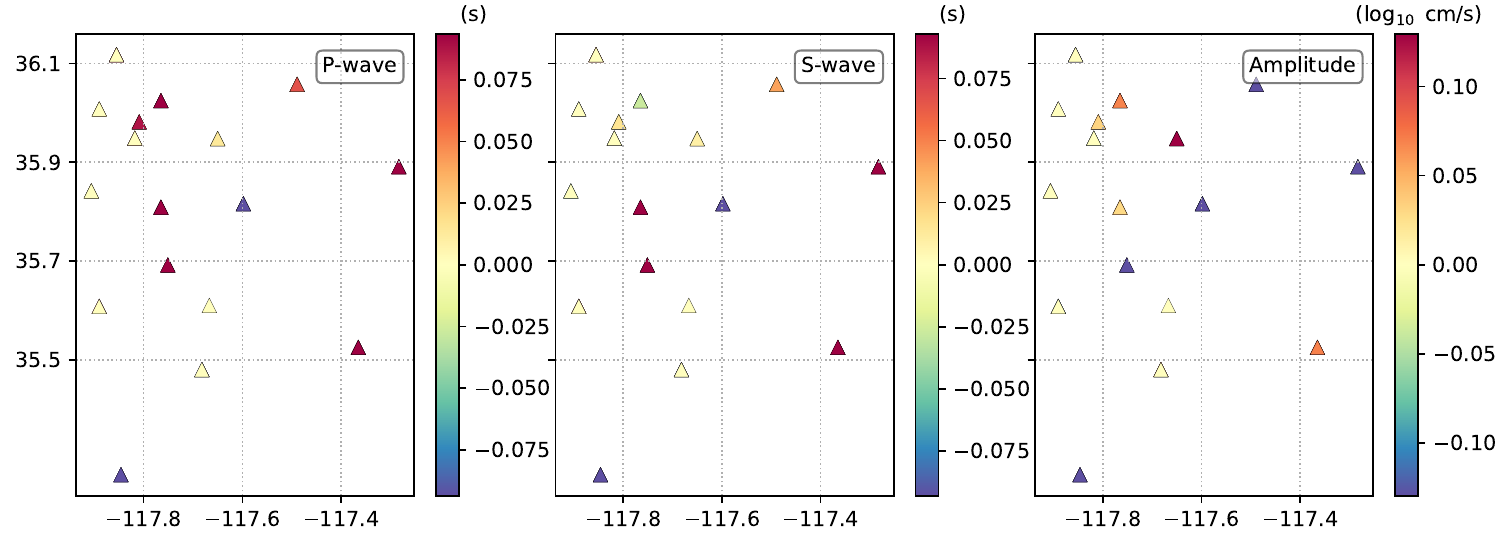}
        \caption{Estimated station correction terms for the Ridgecrest sequence showing (a) P-phase arrival-time corrections, (b) S-phase arrival-time corrections, and (c) amplitude corrections at each station.}
        \label{fig:station_term_ridgecrest}
    \end{figure}

\begin{figure}
        \centering
        \includegraphics[width=0.6\linewidth]{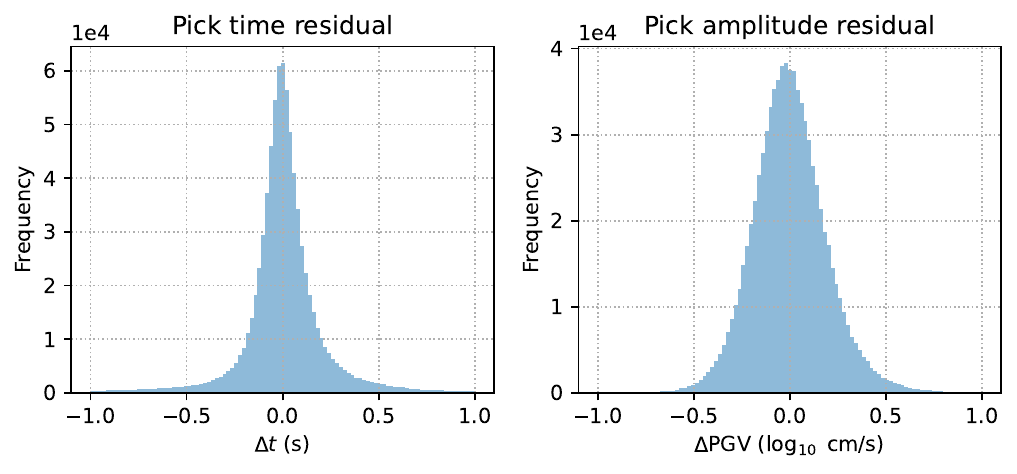}
        \caption{Distribution of pick residuals for (a) P-phase arrival times, (b) S-phase arrival times, and (c) phase amplitudes.}
        \label{fig:pick_residuals}
    \end{figure}

We further evaluated the RANSAC-based earthquake location method on the 2016 Ridgecrest earthquake sequence.
This earthquake sequence is commonly used for evaluating earthquake catalogs due to its extensive aftershock activity and multiple comparison catalogs \citep{ross2019hierarchical,shelly2020high,white2021detailed,liu2020rapid,zhou2023construction,trugman2020stress}.
To facilitate reproducibility, we used the phase picks from \citet{zhu2022earthquake}, generated using a deep-learning-based phase picker, PhaseNet, and a Gaussian mixture model associator, GaMMA.
We analyzed data from 07/04/2019–07/10/2019 within the same spatial extent as \citet{shelly2020higha}'s catalog (-117.8$^\circ$W to -117.3$^\circ$W, 35.5$^\circ$N to 36$^\circ$N).
The dataset includes 21,501 associated pick groups, comprising 452,702 P-phase and 463,217 S-phase picks.
Unlike the synthetic dataset, this real dataset contains phase scores predicted by PhaseNet and measured PGV phase amplitudes, providing additional constraints for outlier detection and enhancing RANSAC robustness.
Phase amplitude information improves outlier identification, particularly for false picks that align with travel-time moveouts but exhibit anomalous amplitudes. Phase scores are incorporated as weights in earthquake location (\Cref{eq:loss}), station term estimation (\Cref{eq:target}), and the minimum pick counts in \Cref{tab:parameters}. Pick counts were determined by summing phase scores rather than raw counts, reducing the influence of low-quality picks.
We applied a simplified ground motion prediction equation for PGV: $\log PGV = -2.175 - 1.68\log R + 0.93 M$ to fit the phase amplitude measurements, where $R$ is hypocentral distance and $M$ is magnitude \citep{picozzi2018rapid}.
Other hyperparameters for defining inliers and outliers are listed in \Cref{tab:parameters}. 
Ten iterations were applied to estimate static station terms for both phase arrival-times and amplitudes. The estimated station correction terms and picks residuals of time and amplitude (\Cref{fig:station_term_ridgecrest,fig:pick_residuals}) show spatial correlation among the three station terms, and demonstrate zero mean, symmetric distributions of pick residuals. 
After RANSAC-based processing, 17,967 events were successfully located using 405,060 P-picks and 414,449 S-picks. 
We compared the catalog with the SCSN catalog and \citet{shelly2020higha}'s catalog, which represent state-of-art in earthquake detection and location. 
Most events in the SCSN catalog have been manually reviewed, ensuring the catalog's high quality during the aftershock period; and \citet{shelly2020higha}'s catalog enhanced magnitude completeness and location accuracy through template matching and double-difference relocation.
Although direct catalog comparisons remain challenging due to the absence of ground truth, \Cref{fig:event_location,fig:event_time,fig:event_mag,} demonstrate consistency in spatial, temporal, and magnitude distributions, validating the effectiveness of ADLoc for complex earthquake sequence data.
The ADLoc locations can be further refined through double-difference earthquake relocation using HypoDD \citep{waldhauser2000doubledifference} and Growclust \citep{trugman2017growclust} in future applications.

\begin{table}[]
\centering
\caption{Hyperparameters for defining inlier picks and valid iterations  in the RANSAC-based location method.}
\label{tab:parameters}
\resizebox{0.4\textwidth}{!}{%
\begin{tabular}{lc}
\hline
Hyperparameter                                                                               & Value \\ \hline
Min picks                                                                                    & 5     \\
Min P-picks                                                                                  & 1     \\
Min S-picks                                                                                  & 1     \\
\begin{tabular}[c]{@{}l@{}}Max time residual (second)\end{tabular}                           & 1     \\
\begin{tabular}[c]{@{}l@{}}Max amplitude residual ($\log_{10}$ cm/s) \end{tabular}           & 1     \\
Min R$^2$-score                                                                              & 0.6   \\ \hline
\end{tabular}%
}
\end{table}

\section{Discussion}

RANSAC is a robust machine learning method designed to handle datasets containing many outliers by leveraging the assumption that a subset of samples can effectively constrain parameters.
This assumption aligns well with earthquake location problems, which involve only four degrees of freedom ($x, y, z, t$) and require a minimum of four measurements.
Through tests on synthetic and real data, we demonstrated the effectiveness of RANSAC-based earthquake location in environments with significant outliers in phase picks.
RANSAC's iterative random sampling and inlier detection enable effectively isolate false picks, reducing their impact on location estimates.
This robustness is particularly valuable in low-magnitude earthquake scenarios where data quality and signal-to-noise ratios are poor, leading to more frequent false or inaccurate picks.

Despite its advantages, the RANSAC-based method has several limitations. The first is increased computational cost due to the random sampling process. The number of RANSAC iterations, $N$, is determined to ensure a high probability ($P$) of selecting at least one all-inlier subset:
\begin{equation}
    N \geq \frac{\log(1 - P)}{\log(1 - R^s)}
\end{equation}
where $P$ is the probability of identifying an all-inlier subset, $R$ is the fraction of inliers in the dataset, and $s$ is the subset size. Common values include $P = 0.99$, with $R$ estimated from the dataset, and $s$ set to the minimum number of picks required for valid earthquake location (\Cref{tab:parameters}).
If no outliers exist, only one iteration is required, and the computational cost reverts to that of traditional earthquake location methods. However, datasets with more outliers necessitate additional iterations and increased computational costs to achieve reliable results. The inlier fraction is dynamically estimated during iterations, and as location accuracy improves, the inlier fraction increases, reducing the maximum required iterations. This iterative refinement allows the algorithm to quickly converge to an all-inlier subset. In practice, a cap on the maximum number of iterations is imposed to control computational expense.

The second limitation is the dependence on hyperparameters for defining outliers. The hyperparameters listed in \Cref{tab:parameters} provide flexibility to optimize location quality but also introduce sensitivity to parameter settings. Synthetic tests (\Cref{tab:synthetic_test}) indicate that tighter thresholds on maximum time residual enable more effective outlier detection, thereby enhancing location accuracy. However, real-world applications require careful tuning of these thresholds to balance the trade-off between the number of successfully located earthquakes and the accuracy of their locations.
When applied to the Ridgecrest sequence, these limitations have minimal impact. The method processes 24,479 events with 1,005,388 picks in approximately 1.5 minutes using 32 cores, and the default hyperparameters yield consistent and reliable location results.

Last, while the RANSAC iterations improve robustness to outliers, they do not guarantee an optimal solution for earthquake location. The core location step, which minimizes travel-time residuals (Step \ref{step:inversion}), remains the same as in traditional earthquake location methods. Therefore, the RANSAC-based framework still inherits limitations of traditional methods, such as potential convergence to local minima, trade-offs between event depth and origin time, and biases from inaccurate velocity models. On the other hand, the RANSAC framework's adaptability allows integration of improvements in both RANSAC and earthquake location algorithms. Our implementation demonstrates this through the successful integration of static station corrections and phase amplitude-based outlier detection.
Moreover, alternative RANSAC extensions, such as MLESAC (Maximum Likelihood Estimation Sample Consensus) \citep{torr2000mlesac}, could enhance the inlier detection process by incorporating probabilistic approaches instead of hard threshold.
Integrating uncertainty estimation frameworks into the RANSAC process could enhance robust earthquake localization in future research.

\section{Conclusions}

Advances in seismic monitoring and machine learning have significantly improved the detection of small-magnitude earthquakes, while introducing new challenges in robust earthquake location due to increased false or inaccurate phase picks.
We have introduced a robust earthquake location framework that integrates the Random Sample Consensus (RANSAC) algorithm with traditional location techniques. The framework enhances robustness against outliers through iterative sampling and outlier detection and is adaptable to further improvements, such as static station corrections and amplitude-based outlier detection.
Through validation with synthetic experiments and application to the Ridgecrest earthquake sequence, the framework has demonstrated consistent performance and enhanced accuracy in outlier-prone scenarios.
The RANSAC-based framework offers a robust and flexible solution to balance the trade-off between expanding earthquake catalogs and maintaining data quality. 
This approach contributes to the development of higher-quality earthquake catalogs, enabling precise earthquake characterization and supporting a wide range of geophysical studies.

\section*{Acknowledgments}

We thank Yifan Yu for sharing the benchmark earthquake location tests \citep{yu2024accuracy} and helping with the calculation of location resolution metrics.
The ADLoc source code is available on GitHub at \href{https://github.com/AI4EPS/ADLoc}{https://github.com/AI4EPS/ADLoc}

\newpage
\clearpage
\bibliography{references}

\begin{thebibliography}{}

\bibitem [\protect \citeauthoryear {%
Aki%
\ \BBA {} Lee%
}{%
Aki%
\ \BBA {} Lee%
}{%
{\protect \APACyear {1976}}%
}]{%
aki1976determination}
\APACinsertmetastar {%
aki1976determination}%
\begin{APACrefauthors}%
Aki, K.%
\BCBT {}\ \BBA {} Lee, W.%
\end{APACrefauthors}%
\unskip\
\newblock
\APACrefYearMonthDay{1976}{}{}.
\newblock
{\BBOQ}\APACrefatitle {Determination of three-dimensional velocity anomalies
  under a seismic array using first P arrival times from local earthquakes: 1.
  A homogeneous initial model} {Determination of three-dimensional velocity
  anomalies under a seismic array using first p arrival times from local
  earthquakes: 1. a homogeneous initial model}.{\BBCQ}
\newblock
\APACjournalVolNumPages{Journal of Geophysical research}{81}{23}{4381--4399}.
\PrintBackRefs{\CurrentBib}

\bibitem [\protect \citeauthoryear {%
Allen%
, Gasparini%
, Kamigaichi%
\BCBL {}\ \BBA {} Bose%
}{%
Allen%
\ \protect \BOthers {.}}{%
{\protect \APACyear {2009}}%
}]{%
allen2009status}
\APACinsertmetastar {%
allen2009status}%
\begin{APACrefauthors}%
Allen, R\BPBI M.%
, Gasparini, P.%
, Kamigaichi, O.%
\BCBL {}\ \BBA {} Bose, M.%
\end{APACrefauthors}%
\unskip\
\newblock
\APACrefYearMonthDay{2009}{}{}.
\newblock
{\BBOQ}\APACrefatitle {The status of earthquake early warning around the world:
  An introductory overview} {The status of earthquake early warning around the
  world: An introductory overview}.{\BBCQ}
\newblock
\APACjournalVolNumPages{Seismological Research Letters}{80}{5}{682--693}.
\PrintBackRefs{\CurrentBib}

\bibitem [\protect \citeauthoryear {%
Anderson%
}{%
Anderson%
}{%
{\protect \APACyear {1982}}%
}]{%
anderson1982robust}
\APACinsertmetastar {%
anderson1982robust}%
\begin{APACrefauthors}%
Anderson, K\BPBI R.%
\end{APACrefauthors}%
\unskip\
\newblock
\APACrefYearMonthDay{1982}{}{}.
\newblock
{\BBOQ}\APACrefatitle {Robust earthquake location using M-estimates} {Robust
  earthquake location using m-estimates}.{\BBCQ}
\newblock
\APACjournalVolNumPages{Physics of the Earth and Planetary
  Interiors}{30}{2-3}{119--130}.
\PrintBackRefs{\CurrentBib}

\bibitem [\protect \citeauthoryear {%
Baydin%
, Pearlmutter%
, Radul%
\BCBL {}\ \BBA {} Siskind%
}{%
Baydin%
\ \protect \BOthers {.}}{%
{\protect \APACyear {2018}}%
}]{%
baydin2018automatic}
\APACinsertmetastar {%
baydin2018automatic}%
\begin{APACrefauthors}%
Baydin, A\BPBI G.%
, Pearlmutter, B\BPBI A.%
, Radul, A\BPBI A.%
\BCBL {}\ \BBA {} Siskind, J\BPBI M.%
\end{APACrefauthors}%
\unskip\
\newblock
\APACrefYearMonthDay{2018}{}{}.
\newblock
{\BBOQ}\APACrefatitle {Automatic differentiation in machine learning: a survey}
  {Automatic differentiation in machine learning: a survey}.{\BBCQ}
\newblock
\APACjournalVolNumPages{Journal of machine learning research}{18}{153}{1--43}.
\PrintBackRefs{\CurrentBib}

\bibitem [\protect \citeauthoryear {%
Byrd%
, Lu%
, Nocedal%
\BCBL {}\ \BBA {} Zhu%
}{%
Byrd%
\ \protect \BOthers {.}}{%
{\protect \APACyear {1995}}%
}]{%
byrd1995limited}
\APACinsertmetastar {%
byrd1995limited}%
\begin{APACrefauthors}%
Byrd, R\BPBI H.%
, Lu, P.%
, Nocedal, J.%
\BCBL {}\ \BBA {} Zhu, C.%
\end{APACrefauthors}%
\unskip\
\newblock
\APACrefYearMonthDay{1995}{}{}.
\newblock
{\BBOQ}\APACrefatitle {A limited memory algorithm for bound constrained
  optimization} {A limited memory algorithm for bound constrained
  optimization}.{\BBCQ}
\newblock
\APACjournalVolNumPages{SIAM Journal on scientific
  computing}{16}{5}{1190--1208}.
\PrintBackRefs{\CurrentBib}

\bibitem [\protect \citeauthoryear {%
{\v{C}}erven{\`y}%
\ \BBA {} P{\v{s}}en{\v{c}}{\'\i}k%
}{%
{\v{C}}erven{\`y}%
\ \BBA {} P{\v{s}}en{\v{c}}{\'\i}k%
}{%
{\protect \APACyear {2020}}%
}]{%
vcerveny2020seismic}
\APACinsertmetastar {%
vcerveny2020seismic}%
\begin{APACrefauthors}%
{\v{C}}erven{\`y}, V.%
\BCBT {}\ \BBA {} P{\v{s}}en{\v{c}}{\'\i}k, I.%
\end{APACrefauthors}%
\unskip\
\newblock
\APACrefYearMonthDay{2020}{}{}.
\newblock
{\BBOQ}\APACrefatitle {Seismic ray theory} {Seismic ray theory}.{\BBCQ}
\newblock
\APACjournalVolNumPages{Encyclopedia of solid earth geophysics}{}{}{1--17}.
\PrintBackRefs{\CurrentBib}

\bibitem [\protect \citeauthoryear {%
Douglas%
}{%
Douglas%
}{%
{\protect \APACyear {1967}}%
}]{%
douglas1967joint}
\APACinsertmetastar {%
douglas1967joint}%
\begin{APACrefauthors}%
Douglas, A.%
\end{APACrefauthors}%
\unskip\
\newblock
\APACrefYearMonthDay{1967}{{\APACmonth{07}}}{}.
\newblock
{\BBOQ}\APACrefatitle {Joint Epicentre Determination} {Joint epicentre
  determination}.{\BBCQ}
\newblock
\APACjournalVolNumPages{Nature}{215}{5096}{47--48}.
\PrintBackRefs{\CurrentBib}

\bibitem [\protect \citeauthoryear {%
Ellsworth%
}{%
Ellsworth%
}{%
{\protect \APACyear {1975}}%
}]{%
ellsworth1975bear}
\APACinsertmetastar {%
ellsworth1975bear}%
\begin{APACrefauthors}%
Ellsworth, W\BPBI L.%
\end{APACrefauthors}%
\unskip\
\newblock
\APACrefYearMonthDay{1975}{}{}.
\newblock
{\BBOQ}\APACrefatitle {Bear Valley, California, Earthquake Sequence of
  {{February-march}} 1972} {Bear valley, california, earthquake sequence of
  {{February-march}} 1972}.{\BBCQ}
\newblock
\APACjournalVolNumPages{Bulletin of the Seismological Society of
  America}{65}{2}{483--506}.
\PrintBackRefs{\CurrentBib}

\bibitem [\protect \citeauthoryear {%
Ellsworth%
}{%
Ellsworth%
}{%
{\protect \APACyear {2013}}%
}]{%
ellsworth2013injection}
\APACinsertmetastar {%
ellsworth2013injection}%
\begin{APACrefauthors}%
Ellsworth, W\BPBI L.%
\end{APACrefauthors}%
\unskip\
\newblock
\APACrefYearMonthDay{2013}{}{}.
\newblock
{\BBOQ}\APACrefatitle {Injection-induced earthquakes} {Injection-induced
  earthquakes}.{\BBCQ}
\newblock
\APACjournalVolNumPages{science}{341}{6142}{1225942}.
\PrintBackRefs{\CurrentBib}

\bibitem [\protect \citeauthoryear {%
Fadakar~Alghalandis%
, Dowd%
\BCBL {}\ \BBA {} Xu%
}{%
Fadakar~Alghalandis%
\ \protect \BOthers {.}}{%
{\protect \APACyear {2013}}%
}]{%
fadakar2013ransac}
\APACinsertmetastar {%
fadakar2013ransac}%
\begin{APACrefauthors}%
Fadakar~Alghalandis, Y.%
, Dowd, P\BPBI A.%
\BCBL {}\ \BBA {} Xu, C.%
\end{APACrefauthors}%
\unskip\
\newblock
\APACrefYearMonthDay{2013}{}{}.
\newblock
{\BBOQ}\APACrefatitle {The {{RANSAC}} Method for Generating Fracture Networks
  from Micro-Seismic Event Data} {The {{RANSAC}} method for generating fracture
  networks from micro-seismic event data}.{\BBCQ}
\newblock
\APACjournalVolNumPages{Mathematical Geosciences}{45}{}{207--224}.
\PrintBackRefs{\CurrentBib}

\bibitem [\protect \citeauthoryear {%
Faulkner%
\ \protect \BOthers {.}}{%
Faulkner%
\ \protect \BOthers {.}}{%
{\protect \APACyear {2010}}%
}]{%
faulkner2010review}
\APACinsertmetastar {%
faulkner2010review}%
\begin{APACrefauthors}%
Faulkner, D.%
, Jackson, C.%
, Lunn, R.%
, Schlische, R.%
, Shipton, Z.%
, Wibberley, C.%
\BCBL {}\ \BBA {} Withjack, M.%
\end{APACrefauthors}%
\unskip\
\newblock
\APACrefYearMonthDay{2010}{}{}.
\newblock
{\BBOQ}\APACrefatitle {A review of recent developments concerning the
  structure, mechanics and fluid flow properties of fault zones} {A review of
  recent developments concerning the structure, mechanics and fluid flow
  properties of fault zones}.{\BBCQ}
\newblock
\APACjournalVolNumPages{Journal of Structural Geology}{32}{11}{1557--1575}.
\PrintBackRefs{\CurrentBib}

\bibitem [\protect \citeauthoryear {%
Fischler%
\ \BBA {} Bolles%
}{%
Fischler%
\ \BBA {} Bolles%
}{%
{\protect \APACyear {1981}}%
}]{%
fischler1981random}
\APACinsertmetastar {%
fischler1981random}%
\begin{APACrefauthors}%
Fischler, M\BPBI A.%
\BCBT {}\ \BBA {} Bolles, R\BPBI C.%
\end{APACrefauthors}%
\unskip\
\newblock
\APACrefYearMonthDay{1981}{}{}.
\newblock
{\BBOQ}\APACrefatitle {Random Sample Consensus: A Paradigm for Model Fitting
  with Applications to Image Analysis and Automated Cartography} {Random sample
  consensus: A paradigm for model fitting with applications to image analysis
  and automated cartography}.{\BBCQ}
\newblock
\APACjournalVolNumPages{Communications of the ACM}{24}{6}{381--395}.
\PrintBackRefs{\CurrentBib}

\bibitem [\protect \citeauthoryear {%
Font%
, Kao%
, Lallemand%
, Liu%
\BCBL {}\ \BBA {} Chiao%
}{%
Font%
\ \protect \BOthers {.}}{%
{\protect \APACyear {2004}}%
}]{%
font2004hypocentre}
\APACinsertmetastar {%
font2004hypocentre}%
\begin{APACrefauthors}%
Font, Y.%
, Kao, H.%
, Lallemand, S.%
, Liu, C\BHBI S.%
\BCBL {}\ \BBA {} Chiao, L\BHBI Y.%
\end{APACrefauthors}%
\unskip\
\newblock
\APACrefYearMonthDay{2004}{}{}.
\newblock
{\BBOQ}\APACrefatitle {Hypocentre determination offshore of eastern Taiwan
  using the Maximum Intersection method} {Hypocentre determination offshore of
  eastern taiwan using the maximum intersection method}.{\BBCQ}
\newblock
\APACjournalVolNumPages{Geophysical Journal International}{158}{2}{655--675}.
\PrintBackRefs{\CurrentBib}

\bibitem [\protect \citeauthoryear {%
Frohlich%
}{%
Frohlich%
}{%
{\protect \APACyear {1979}}%
}]{%
frohlich1979efficient}
\APACinsertmetastar {%
frohlich1979efficient}%
\begin{APACrefauthors}%
Frohlich, C.%
\end{APACrefauthors}%
\unskip\
\newblock
\APACrefYearMonthDay{1979}{}{}.
\newblock
{\BBOQ}\APACrefatitle {An efficient method for joint hypocenter determination
  for large groups of earthquakes} {An efficient method for joint hypocenter
  determination for large groups of earthquakes}.{\BBCQ}
\newblock
\APACjournalVolNumPages{Computers \& Geosciences}{5}{3-4}{387--389}.
\PrintBackRefs{\CurrentBib}

\bibitem [\protect \citeauthoryear {%
Geiger%
}{%
Geiger%
}{%
{\protect \APACyear {1912}}%
}]{%
geiger1912probability}
\APACinsertmetastar {%
geiger1912probability}%
\begin{APACrefauthors}%
Geiger, L.%
\end{APACrefauthors}%
\unskip\
\newblock
\APACrefYearMonthDay{1912}{}{}.
\newblock
{\BBOQ}\APACrefatitle {Probability Method for the Determination of Earthquake
  Epicentres from the Arrival Time Only} {Probability method for the
  determination of earthquake epicentres from the arrival time only}.{\BBCQ}
\newblock
\APACjournalVolNumPages{Bull. St. Louis Univ.}{8}{}{60}.
\PrintBackRefs{\CurrentBib}

\bibitem [\protect \citeauthoryear {%
Hartley%
\ \BBA {} Zisserman%
}{%
Hartley%
\ \BBA {} Zisserman%
}{%
{\protect \APACyear {2003}}%
}]{%
hartley2003multiple}
\APACinsertmetastar {%
hartley2003multiple}%
\begin{APACrefauthors}%
Hartley, R.%
\BCBT {}\ \BBA {} Zisserman, A.%
\end{APACrefauthors}%
\unskip\
\newblock
\APACrefYear{2003}.
\newblock
\APACrefbtitle {Multiple View Geometry in Computer Vision} {Multiple view
  geometry in computer vision}.
\newblock
\APACaddressPublisher{}{Cambridge university press}.
\PrintBackRefs{\CurrentBib}

\bibitem [\protect \citeauthoryear {%
Hauksson%
, Yang%
\BCBL {}\ \BBA {} Shearer%
}{%
Hauksson%
\ \protect \BOthers {.}}{%
{\protect \APACyear {2012}}%
}]{%
hauksson2012waveform}
\APACinsertmetastar {%
hauksson2012waveform}%
\begin{APACrefauthors}%
Hauksson, E.%
, Yang, W.%
\BCBL {}\ \BBA {} Shearer, P\BPBI M.%
\end{APACrefauthors}%
\unskip\
\newblock
\APACrefYearMonthDay{2012}{}{}.
\newblock
{\BBOQ}\APACrefatitle {Waveform relocated earthquake catalog for southern
  California (1981 to June 2011)} {Waveform relocated earthquake catalog for
  southern california (1981 to june 2011)}.{\BBCQ}
\newblock
\APACjournalVolNumPages{Bulletin of the Seismological Society of
  America}{102}{5}{2239--2244}.
\PrintBackRefs{\CurrentBib}

\bibitem [\protect \citeauthoryear {%
Hirata%
\ \BBA {} Matsu'ura%
}{%
Hirata%
\ \BBA {} Matsu'ura%
}{%
{\protect \APACyear {1987}}%
}]{%
hirata1987maximumlikelihood}
\APACinsertmetastar {%
hirata1987maximumlikelihood}%
\begin{APACrefauthors}%
Hirata, N.%
\BCBT {}\ \BBA {} Matsu'ura, M.%
\end{APACrefauthors}%
\unskip\
\newblock
\APACrefYearMonthDay{1987}{{\APACmonth{08}}}{}.
\newblock
{\BBOQ}\APACrefatitle {Maximum-Likelihood Estimation of Hypocenter with Origin
  Time Eliminated Using Nonlinear Inversion Technique} {Maximum-likelihood
  estimation of hypocenter with origin time eliminated using nonlinear
  inversion technique}.{\BBCQ}
\newblock
\APACjournalVolNumPages{Physics of the Earth and Planetary
  Interiors}{47}{}{50--61}.
\PrintBackRefs{\CurrentBib}

\bibitem [\protect \citeauthoryear {%
Huber%
}{%
Huber%
}{%
{\protect \APACyear {1964}}%
}]{%
huber1964robust}
\APACinsertmetastar {%
huber1964robust}%
\begin{APACrefauthors}%
Huber, P\BPBI J.%
\end{APACrefauthors}%
\unskip\
\newblock
\APACrefYearMonthDay{1964}{}{}.
\newblock
{\BBOQ}\APACrefatitle {Robust Estimation of a Location Parameter} {Robust
  estimation of a location parameter}.{\BBCQ}
\newblock
\APACjournalVolNumPages{The Annals of Mathematical Statistics}{35}{1}{73--101}.
\PrintBackRefs{\CurrentBib}

\bibitem [\protect \citeauthoryear {%
Hutton%
, Woessner%
\BCBL {}\ \BBA {} Hauksson%
}{%
Hutton%
\ \protect \BOthers {.}}{%
{\protect \APACyear {2010}}%
}]{%
hutton2010earthquake}
\APACinsertmetastar {%
hutton2010earthquake}%
\begin{APACrefauthors}%
Hutton, K.%
, Woessner, J.%
\BCBL {}\ \BBA {} Hauksson, E.%
\end{APACrefauthors}%
\unskip\
\newblock
\APACrefYearMonthDay{2010}{{\APACmonth{04}}}{}.
\newblock
{\BBOQ}\APACrefatitle {Earthquake {{Monitoring}} in {{Southern California}} for
  {{Seventy-Seven Years}} (1932-2008)} {Earthquake {{Monitoring}} in {{Southern
  California}} for {{Seventy-Seven Years}} (1932-2008)}.{\BBCQ}
\newblock
\APACjournalVolNumPages{Bulletin of the Seismological Society of
  America}{100}{2}{423--446}.
\PrintBackRefs{\CurrentBib}

\bibitem [\protect \citeauthoryear {%
Kanamori%
}{%
Kanamori%
}{%
{\protect \APACyear {2005}}%
}]{%
kanamori2005real}
\APACinsertmetastar {%
kanamori2005real}%
\begin{APACrefauthors}%
Kanamori, H.%
\end{APACrefauthors}%
\unskip\
\newblock
\APACrefYearMonthDay{2005}{}{}.
\newblock
{\BBOQ}\APACrefatitle {Real-time seismology and earthquake damage mitigation}
  {Real-time seismology and earthquake damage mitigation}.{\BBCQ}
\newblock
\APACjournalVolNumPages{Annu. Rev. Earth Planet. Sci.}{33}{1}{195--214}.
\PrintBackRefs{\CurrentBib}

\bibitem [\protect \citeauthoryear {%
Kanamori%
\ \BBA {} Brodsky%
}{%
Kanamori%
\ \BBA {} Brodsky%
}{%
{\protect \APACyear {2004}}%
}]{%
kanamori2004physics}
\APACinsertmetastar {%
kanamori2004physics}%
\begin{APACrefauthors}%
Kanamori, H.%
\BCBT {}\ \BBA {} Brodsky, E\BPBI E.%
\end{APACrefauthors}%
\unskip\
\newblock
\APACrefYearMonthDay{2004}{}{}.
\newblock
{\BBOQ}\APACrefatitle {The physics of earthquakes} {The physics of
  earthquakes}.{\BBCQ}
\newblock
\APACjournalVolNumPages{Reports on progress in physics}{67}{8}{1429}.
\PrintBackRefs{\CurrentBib}

\bibitem [\protect \citeauthoryear {%
Kissling%
, Ellsworth%
, {Eberhart-Phillips}%
\BCBL {}\ \BBA {} Kradolfer%
}{%
Kissling%
\ \protect \BOthers {.}}{%
{\protect \APACyear {1994}}%
}]{%
kissling1994initial}
\APACinsertmetastar {%
kissling1994initial}%
\begin{APACrefauthors}%
Kissling, E.%
, Ellsworth, W\BPBI L.%
, {Eberhart-Phillips}, D.%
\BCBL {}\ \BBA {} Kradolfer, U.%
\end{APACrefauthors}%
\unskip\
\newblock
\APACrefYearMonthDay{1994}{}{}.
\newblock
{\BBOQ}\APACrefatitle {Initial Reference Models in Local Earthquake Tomography}
  {Initial reference models in local earthquake tomography}.{\BBCQ}
\newblock
\APACjournalVolNumPages{Journal of Geophysical Research: Solid
  Earth}{99}{B10}{19635--19646}.
\PrintBackRefs{\CurrentBib}

\bibitem [\protect \citeauthoryear {%
Klein%
}{%
Klein%
}{%
{\protect \APACyear {2002}}%
}]{%
klein2002users}
\APACinsertmetastar {%
klein2002users}%
\begin{APACrefauthors}%
Klein, F\BPBI W.%
\end{APACrefauthors}%
\unskip\
\newblock
\APACrefYearMonthDay{2002}{}{}.
\newblock
\APACrefbtitle {User's Guide to {{HYPOINVERSE-2000}}, a Fortran Program to
  Solve for Earthquake Locations and Magnitudes} {User's guide to
  {{HYPOINVERSE-2000}}, a fortran program to solve for earthquake locations and
  magnitudes}\ \APACbVolEdTR{}{\BTR{}\ \BNUM\ 2002-171}.
\newblock
\APACaddressInstitution{}{U.S. Geological Survey}.
\PrintBackRefs{\CurrentBib}

\bibitem [\protect \citeauthoryear {%
Lahr%
}{%
Lahr%
}{%
{\protect \APACyear {1979}}%
}]{%
lahr1979hypoellipse}
\APACinsertmetastar {%
lahr1979hypoellipse}%
\begin{APACrefauthors}%
Lahr, J\BPBI C.%
\end{APACrefauthors}%
\unskip\
\newblock
\APACrefYearMonthDay{1979}{}{}.
\newblock
\APACrefbtitle {{{HYPOELLIPSE}}/{{MULTICS}}: A Computer Program for Determining
  Local Earthquake Hypocentral Parameters, Magnitude, and First Motion Pattern}
  {{{HYPOELLIPSE}}/{{MULTICS}}: A computer program for determining local
  earthquake hypocentral parameters, magnitude, and first motion pattern}\
  \APACbVolEdTR{}{\BTR{}}.
\newblock
\APACaddressInstitution{}{US Geological Survey}.
\PrintBackRefs{\CurrentBib}

\bibitem [\protect \citeauthoryear {%
Lee%
\ \BBA {} Lahr%
}{%
Lee%
\ \BBA {} Lahr%
}{%
{\protect \APACyear {1975}}%
}]{%
lee1975hypo71}
\APACinsertmetastar {%
lee1975hypo71}%
\begin{APACrefauthors}%
Lee, W\BPBI H.%
\BCBT {}\ \BBA {} Lahr, J\BPBI C.%
\end{APACrefauthors}%
\unskip\
\newblock
\APACrefYearMonthDay{1975}{}{}.
\newblock
\APACrefbtitle {{{HYPO71}} (Revised; a Computer Program for Determining
  Hypocenter, Magnitude, and First Motion Pattern of Local Earthquakes}
  {{{HYPO71}} (revised; a computer program for determining hypocenter,
  magnitude, and first motion pattern of local earthquakes}\
  \APACbVolEdTR{}{\BTR{}}.
\newblock
\APACaddressInstitution{}{US Dept. of the Interior, Geological Survey, National
  Center for Earthquake {\dots}}.
\PrintBackRefs{\CurrentBib}

\bibitem [\protect \citeauthoryear {%
Lin%
}{%
Lin%
}{%
{\protect \APACyear {2018}}%
}]{%
lin2018source}
\APACinsertmetastar {%
lin2018source}%
\begin{APACrefauthors}%
Lin, G.%
\end{APACrefauthors}%
\unskip\
\newblock
\APACrefYearMonthDay{2018}{{\APACmonth{09}}}{}.
\newblock
{\BBOQ}\APACrefatitle {The {{Source}}-{{Specific Station Term}} and {{Waveform
  Cross}}-{{Correlation Earthquake Location Package}} and {{Its Applications}}
  to {{California}} and {{New Zealand}}} {The {{Source}}-{{Specific Station
  Term}} and {{Waveform Cross}}-{{Correlation Earthquake Location Package}} and
  {{Its Applications}} to {{California}} and {{New Zealand}}}.{\BBCQ}
\newblock
\APACjournalVolNumPages{Seismological Research Letters}{89}{5}{1877--1885}.
\PrintBackRefs{\CurrentBib}

\bibitem [\protect \citeauthoryear {%
Lin%
\ \BBA {} Shearer%
}{%
Lin%
\ \BBA {} Shearer%
}{%
{\protect \APACyear {2006}}%
}]{%
lin2006comploca}
\APACinsertmetastar {%
lin2006comploca}%
\begin{APACrefauthors}%
Lin, G.%
\BCBT {}\ \BBA {} Shearer, P.%
\end{APACrefauthors}%
\unskip\
\newblock
\APACrefYearMonthDay{2006}{{\APACmonth{07}}}{}.
\newblock
{\BBOQ}\APACrefatitle {The {{COMPLOC Earthquake Location Package}}} {The
  {{COMPLOC Earthquake Location Package}}}.{\BBCQ}
\newblock
\APACjournalVolNumPages{Seismological Research Letters}{77}{4}{440--444}.
\PrintBackRefs{\CurrentBib}

\bibitem [\protect \citeauthoryear {%
M.~Liu%
, Zhang%
, Zhu%
, Ellsworth%
\BCBL {}\ \BBA {} Li%
}{%
M.~Liu%
\ \protect \BOthers {.}}{%
{\protect \APACyear {2020}}%
}]{%
liu2020rapid}
\APACinsertmetastar {%
liu2020rapid}%
\begin{APACrefauthors}%
Liu, M.%
, Zhang, M.%
, Zhu, W.%
, Ellsworth, W\BPBI L.%
\BCBL {}\ \BBA {} Li, H.%
\end{APACrefauthors}%
\unskip\
\newblock
\APACrefYearMonthDay{2020}{}{}.
\newblock
{\BBOQ}\APACrefatitle {Rapid characterization of the July 2019 Ridgecrest,
  California, earthquake sequence from raw seismic data using machine-learning
  phase picker} {Rapid characterization of the july 2019 ridgecrest,
  california, earthquake sequence from raw seismic data using machine-learning
  phase picker}.{\BBCQ}
\newblock
\APACjournalVolNumPages{Geophysical Research Letters}{47}{4}{e2019GL086189}.
\PrintBackRefs{\CurrentBib}

\bibitem [\protect \citeauthoryear {%
X.~Liu%
, Jin%
, Lin%
, Xiang%
\BCBL {}\ \BBA {} Zhong%
}{%
X.~Liu%
\ \protect \BOthers {.}}{%
{\protect \APACyear {2018}}%
}]{%
liu2018enhanced}
\APACinsertmetastar {%
liu2018enhanced}%
\begin{APACrefauthors}%
Liu, X.%
, Jin, Y.%
, Lin, B.%
, Xiang, J.%
\BCBL {}\ \BBA {} Zhong, H.%
\end{APACrefauthors}%
\unskip\
\newblock
\APACrefYearMonthDay{2018}{}{}.
\newblock
{\BBOQ}\APACrefatitle {An Enhanced {{RANSAC}} Method for Complex Hydraulic
  Network Characterization Based on Microseismic Data} {An enhanced {{RANSAC}}
  method for complex hydraulic network characterization based on microseismic
  data}.{\BBCQ}
\newblock
\BIn{} \APACrefbtitle {{{ARMA}} International Discrete Fracture Network
  Engineering Conference} {{{ARMA}} international discrete fracture network
  engineering conference}\ (\BPG~D023S014R010).
\newblock
\APACaddressPublisher{}{ARMA}.
\PrintBackRefs{\CurrentBib}

\bibitem [\protect \citeauthoryear {%
Lomax%
}{%
Lomax%
}{%
{\protect \APACyear {2005}}%
}]{%
lomax2005reanalysis}
\APACinsertmetastar {%
lomax2005reanalysis}%
\begin{APACrefauthors}%
Lomax, A.%
\end{APACrefauthors}%
\unskip\
\newblock
\APACrefYearMonthDay{2005}{}{}.
\newblock
{\BBOQ}\APACrefatitle {A reanalysis of the hypocentral location and related
  observations for the great 1906 California earthquake} {A reanalysis of the
  hypocentral location and related observations for the great 1906 california
  earthquake}.{\BBCQ}
\newblock
\APACjournalVolNumPages{Bulletin of the Seismological Society of
  America}{95}{3}{861--877}.
\PrintBackRefs{\CurrentBib}

\bibitem [\protect \citeauthoryear {%
Lomax%
, Michelini%
\BCBL {}\ \BBA {} Curtis%
}{%
Lomax%
\ \protect \BOthers {.}}{%
{\protect \APACyear {2009}}%
}]{%
lomax2009earthquake}
\APACinsertmetastar {%
lomax2009earthquake}%
\begin{APACrefauthors}%
Lomax, A.%
, Michelini, A.%
\BCBL {}\ \BBA {} Curtis, A.%
\end{APACrefauthors}%
\unskip\
\newblock
\APACrefYearMonthDay{2009}{}{}.
\newblock
{\BBOQ}\APACrefatitle {Earthquake Location, Direct, Global-Search Methods}
  {Earthquake location, direct, global-search methods}.{\BBCQ}
\newblock
\APACjournalVolNumPages{}{}{}{25}.
\PrintBackRefs{\CurrentBib}

\bibitem [\protect \citeauthoryear {%
Lomax%
\ \BBA {} Savvaidis%
}{%
Lomax%
\ \BBA {} Savvaidis%
}{%
{\protect \APACyear {2022}}%
}]{%
lomax2022high}
\APACinsertmetastar {%
lomax2022high}%
\begin{APACrefauthors}%
Lomax, A.%
\BCBT {}\ \BBA {} Savvaidis, A.%
\end{APACrefauthors}%
\unskip\
\newblock
\APACrefYearMonthDay{2022}{}{}.
\newblock
{\BBOQ}\APACrefatitle {High-precision earthquake location using source-specific
  station terms and inter-event waveform similarity} {High-precision earthquake
  location using source-specific station terms and inter-event waveform
  similarity}.{\BBCQ}
\newblock
\APACjournalVolNumPages{Journal of Geophysical Research: Solid
  Earth}{127}{1}{e2021JB023190}.
\PrintBackRefs{\CurrentBib}

\bibitem [\protect \citeauthoryear {%
Lomax%
, Virieux%
, Volant%
\BCBL {}\ \BBA {} {Berge-Thierry}%
}{%
Lomax%
\ \protect \BOthers {.}}{%
{\protect \APACyear {2000}}%
}]{%
lomax2000probabilistic}
\APACinsertmetastar {%
lomax2000probabilistic}%
\begin{APACrefauthors}%
Lomax, A.%
, Virieux, J.%
, Volant, P.%
\BCBL {}\ \BBA {} {Berge-Thierry}, C.%
\end{APACrefauthors}%
\unskip\
\newblock
\APACrefYearMonthDay{2000}{}{}.
\newblock
{\BBOQ}\APACrefatitle {Probabilistic Earthquake Location in {{3D}} and Layered
  Models} {Probabilistic earthquake location in {{3D}} and layered
  models}.{\BBCQ}
\newblock
\BIn{} C\BPBI H.~Thurber\ \BBA {} N.~Rabinowitz\ (\BEDS), \APACrefbtitle
  {Advances in {{Seismic Event Location}}} {Advances in {{Seismic Event
  Location}}}\ (\BPGS\ 101--134).
\newblock
\APACaddressPublisher{Dordrecht}{Springer Netherlands}.
\PrintBackRefs{\CurrentBib}

\bibitem [\protect \citeauthoryear {%
Mousavi%
, Ellsworth%
, Zhu%
, Chuang%
\BCBL {}\ \BBA {} Beroza%
}{%
Mousavi%
\ \protect \BOthers {.}}{%
{\protect \APACyear {2020}}%
}]{%
mousavi2020earthquake}
\APACinsertmetastar {%
mousavi2020earthquake}%
\begin{APACrefauthors}%
Mousavi, S\BPBI M.%
, Ellsworth, W\BPBI L.%
, Zhu, W.%
, Chuang, L\BPBI Y.%
\BCBL {}\ \BBA {} Beroza, G\BPBI C.%
\end{APACrefauthors}%
\unskip\
\newblock
\APACrefYearMonthDay{2020}{}{}.
\newblock
{\BBOQ}\APACrefatitle {Earthquake transformer—an attentive deep-learning
  model for simultaneous earthquake detection and phase picking} {Earthquake
  transformer—an attentive deep-learning model for simultaneous earthquake
  detection and phase picking}.{\BBCQ}
\newblock
\APACjournalVolNumPages{Nature communications}{11}{1}{3952}.
\PrintBackRefs{\CurrentBib}

\bibitem [\protect \citeauthoryear {%
Myers%
, Johannesson%
\BCBL {}\ \BBA {} Hanley%
}{%
Myers%
\ \protect \BOthers {.}}{%
{\protect \APACyear {2007}}%
}]{%
myers2007bayesian}
\APACinsertmetastar {%
myers2007bayesian}%
\begin{APACrefauthors}%
Myers, S\BPBI C.%
, Johannesson, G.%
\BCBL {}\ \BBA {} Hanley, W.%
\end{APACrefauthors}%
\unskip\
\newblock
\APACrefYearMonthDay{2007}{}{}.
\newblock
{\BBOQ}\APACrefatitle {A Bayesian hierarchical method for multiple-event
  seismic location} {A bayesian hierarchical method for multiple-event seismic
  location}.{\BBCQ}
\newblock
\APACjournalVolNumPages{Geophysical Journal International}{171}{3}{1049--1063}.
\PrintBackRefs{\CurrentBib}

\bibitem [\protect \citeauthoryear {%
Nocedal%
\ \BBA {} Wright%
}{%
Nocedal%
\ \BBA {} Wright%
}{%
{\protect \APACyear {1999}}%
}]{%
nocedal1999numerical}
\APACinsertmetastar {%
nocedal1999numerical}%
\begin{APACrefauthors}%
Nocedal, J.%
\BCBT {}\ \BBA {} Wright, S\BPBI J.%
\end{APACrefauthors}%
\unskip\
\newblock
\APACrefYear{1999}.
\newblock
\APACrefbtitle {Numerical optimization} {Numerical optimization}.
\newblock
\APACaddressPublisher{}{Springer}.
\PrintBackRefs{\CurrentBib}

\bibitem [\protect \citeauthoryear {%
Park%
, Beroza%
\BCBL {}\ \BBA {} Ellsworth%
}{%
Park%
\ \protect \BOthers {.}}{%
{\protect \APACyear {2022}}%
}]{%
park2022basement}
\APACinsertmetastar {%
park2022basement}%
\begin{APACrefauthors}%
Park, Y.%
, Beroza, G\BPBI C.%
\BCBL {}\ \BBA {} Ellsworth, W\BPBI L.%
\end{APACrefauthors}%
\unskip\
\newblock
\APACrefYearMonthDay{2022}{}{}.
\newblock
{\BBOQ}\APACrefatitle {Basement fault activation before larger earthquakes in
  Oklahoma and Kansas} {Basement fault activation before larger earthquakes in
  oklahoma and kansas}.{\BBCQ}
\newblock
\APACjournalVolNumPages{The Seismic Record}{2}{3}{197--206}.
\PrintBackRefs{\CurrentBib}

\bibitem [\protect \citeauthoryear {%
Paszke%
\ \protect \BOthers {.}}{%
Paszke%
\ \protect \BOthers {.}}{%
{\protect \APACyear {2017}}%
}]{%
paszke2017automatic}
\APACinsertmetastar {%
paszke2017automatic}%
\begin{APACrefauthors}%
Paszke, A.%
, Gross, S.%
, Chintala, S.%
, Chanan, G.%
, Yang, E.%
, DeVito, Z.%
\BDBL {}Lerer, A.%
\end{APACrefauthors}%
\unskip\
\newblock
\APACrefYearMonthDay{2017}{}{}.
\newblock
{\BBOQ}\APACrefatitle {Automatic differentiation in pytorch} {Automatic
  differentiation in pytorch}.{\BBCQ}
\newblock

\PrintBackRefs{\CurrentBib}

\bibitem [\protect \citeauthoryear {%
Picozzi%
, Bindi%
, Spallarossa%
, Di~Giacomo%
\BCBL {}\ \BBA {} Zollo%
}{%
Picozzi%
\ \protect \BOthers {.}}{%
{\protect \APACyear {2018}}%
}]{%
picozzi2018rapid}
\APACinsertmetastar {%
picozzi2018rapid}%
\begin{APACrefauthors}%
Picozzi, M.%
, Bindi, D.%
, Spallarossa, D.%
, Di~Giacomo, D.%
\BCBL {}\ \BBA {} Zollo, A.%
\end{APACrefauthors}%
\unskip\
\newblock
\APACrefYearMonthDay{2018}{}{}.
\newblock
{\BBOQ}\APACrefatitle {A rapid response magnitude scale for timely assessment
  of the high frequency seismic radiation} {A rapid response magnitude scale
  for timely assessment of the high frequency seismic radiation}.{\BBCQ}
\newblock
\APACjournalVolNumPages{Scientific reports}{8}{1}{8562}.
\PrintBackRefs{\CurrentBib}

\bibitem [\protect \citeauthoryear {%
Podvin%
\ \BBA {} Lecomte%
}{%
Podvin%
\ \BBA {} Lecomte%
}{%
{\protect \APACyear {1991}}%
}]{%
podvin1991finite}
\APACinsertmetastar {%
podvin1991finite}%
\begin{APACrefauthors}%
Podvin, P.%
\BCBT {}\ \BBA {} Lecomte, I.%
\end{APACrefauthors}%
\unskip\
\newblock
\APACrefYearMonthDay{1991}{}{}.
\newblock
{\BBOQ}\APACrefatitle {Finite difference computation of traveltimes in very
  contrasted velocity models: a massively parallel approach and its associated
  tools} {Finite difference computation of traveltimes in very contrasted
  velocity models: a massively parallel approach and its associated
  tools}.{\BBCQ}
\newblock
\APACjournalVolNumPages{Geophysical Journal International}{105}{1}{271--284}.
\PrintBackRefs{\CurrentBib}

\bibitem [\protect \citeauthoryear {%
Pollitz%
, Wicks%
, Schoenball%
, Ellsworth%
\BCBL {}\ \BBA {} Murray%
}{%
Pollitz%
\ \protect \BOthers {.}}{%
{\protect \APACyear {2017}}%
}]{%
pollitz2017geodetic}
\APACinsertmetastar {%
pollitz2017geodetic}%
\begin{APACrefauthors}%
Pollitz, F\BPBI F.%
, Wicks, C.%
, Schoenball, M.%
, Ellsworth, W.%
\BCBL {}\ \BBA {} Murray, M.%
\end{APACrefauthors}%
\unskip\
\newblock
\APACrefYearMonthDay{2017}{}{}.
\newblock
{\BBOQ}\APACrefatitle {Geodetic Slip Model of the 3 September 2016 {{M}} w 5.8
  Pawnee, Oklahoma, Earthquake: Evidence for Fault-Zone Collapse} {Geodetic
  slip model of the 3 september 2016 {{M}} w 5.8 pawnee, oklahoma, earthquake:
  Evidence for fault-zone collapse}.{\BBCQ}
\newblock
\APACjournalVolNumPages{Seismological Research Letters}{88}{4}{983--993}.
\PrintBackRefs{\CurrentBib}

\bibitem [\protect \citeauthoryear {%
Pujol%
}{%
Pujol%
}{%
{\protect \APACyear {1988}}%
}]{%
pujol1988comments}
\APACinsertmetastar {%
pujol1988comments}%
\begin{APACrefauthors}%
Pujol, J.%
\end{APACrefauthors}%
\unskip\
\newblock
\APACrefYearMonthDay{1988}{}{}.
\newblock
{\BBOQ}\APACrefatitle {Comments on the Joint Determination of Hypocenters and
  Station Corrections} {Comments on the joint determination of hypocenters and
  station corrections}.{\BBCQ}
\newblock
\APACjournalVolNumPages{Bulletin of the Seismological Society of
  America}{78}{3}{1179--1189}.
\PrintBackRefs{\CurrentBib}

\bibitem [\protect \citeauthoryear {%
Richards-Dinger%
\ \BBA {} Shearer%
}{%
Richards-Dinger%
\ \BBA {} Shearer%
}{%
{\protect \APACyear {2000}}%
}]{%
richards2000earthquake}
\APACinsertmetastar {%
richards2000earthquake}%
\begin{APACrefauthors}%
Richards-Dinger, K.%
\BCBT {}\ \BBA {} Shearer, P.%
\end{APACrefauthors}%
\unskip\
\newblock
\APACrefYearMonthDay{2000}{}{}.
\newblock
{\BBOQ}\APACrefatitle {Earthquake locations in southern California obtained
  using source-specific station terms} {Earthquake locations in southern
  california obtained using source-specific station terms}.{\BBCQ}
\newblock
\APACjournalVolNumPages{Journal of Geophysical Research: Solid
  Earth}{105}{B5}{10939--10960}.
\PrintBackRefs{\CurrentBib}

\bibitem [\protect \citeauthoryear {%
Ross%
, Cochran%
, Trugman%
\BCBL {}\ \BBA {} Smith%
}{%
Ross%
\ \protect \BOthers {.}}{%
{\protect \APACyear {2020}}%
}]{%
ross20203d}
\APACinsertmetastar {%
ross20203d}%
\begin{APACrefauthors}%
Ross, Z\BPBI E.%
, Cochran, E\BPBI S.%
, Trugman, D\BPBI T.%
\BCBL {}\ \BBA {} Smith, J\BPBI D.%
\end{APACrefauthors}%
\unskip\
\newblock
\APACrefYearMonthDay{2020}{}{}.
\newblock
{\BBOQ}\APACrefatitle {3D fault architecture controls the dynamism of
  earthquake swarms} {3d fault architecture controls the dynamism of earthquake
  swarms}.{\BBCQ}
\newblock
\APACjournalVolNumPages{Science}{368}{6497}{1357--1361}.
\PrintBackRefs{\CurrentBib}

\bibitem [\protect \citeauthoryear {%
Ross%
\ \protect \BOthers {.}}{%
Ross%
\ \protect \BOthers {.}}{%
{\protect \APACyear {2019}}%
}]{%
ross2019hierarchical}
\APACinsertmetastar {%
ross2019hierarchical}%
\begin{APACrefauthors}%
Ross, Z\BPBI E.%
, Idini, B.%
, Jia, Z.%
, Stephenson, O\BPBI L.%
, Zhong, M.%
, Wang, X.%
\BDBL {}others%
\end{APACrefauthors}%
\unskip\
\newblock
\APACrefYearMonthDay{2019}{}{}.
\newblock
{\BBOQ}\APACrefatitle {Hierarchical interlocked orthogonal faulting in the 2019
  Ridgecrest earthquake sequence} {Hierarchical interlocked orthogonal faulting
  in the 2019 ridgecrest earthquake sequence}.{\BBCQ}
\newblock
\APACjournalVolNumPages{Science}{366}{6463}{346--351}.
\PrintBackRefs{\CurrentBib}

\bibitem [\protect \citeauthoryear {%
Ross%
, Meier%
\BCBL {}\ \BBA {} Hauksson%
}{%
Ross%
\ \protect \BOthers {.}}{%
{\protect \APACyear {2018}}%
}]{%
ross2018p}
\APACinsertmetastar {%
ross2018p}%
\begin{APACrefauthors}%
Ross, Z\BPBI E.%
, Meier, M\BHBI A.%
\BCBL {}\ \BBA {} Hauksson, E.%
\end{APACrefauthors}%
\unskip\
\newblock
\APACrefYearMonthDay{2018}{}{}.
\newblock
{\BBOQ}\APACrefatitle {P wave arrival picking and first-motion polarity
  determination with deep learning} {P wave arrival picking and first-motion
  polarity determination with deep learning}.{\BBCQ}
\newblock
\APACjournalVolNumPages{Journal of Geophysical Research: Solid
  Earth}{123}{6}{5120--5129}.
\PrintBackRefs{\CurrentBib}

\bibitem [\protect \citeauthoryear {%
P.~Shearer%
, Hauksson%
\BCBL {}\ \BBA {} Lin%
}{%
P.~Shearer%
\ \protect \BOthers {.}}{%
{\protect \APACyear {2005}}%
}]{%
shearer2005southern}
\APACinsertmetastar {%
shearer2005southern}%
\begin{APACrefauthors}%
Shearer, P.%
, Hauksson, E.%
\BCBL {}\ \BBA {} Lin, G.%
\end{APACrefauthors}%
\unskip\
\newblock
\APACrefYearMonthDay{2005}{}{}.
\newblock
{\BBOQ}\APACrefatitle {Southern California hypocenter relocation with waveform
  cross-correlation, Part 2: Results using source-specific station terms and
  cluster analysis} {Southern california hypocenter relocation with waveform
  cross-correlation, part 2: Results using source-specific station terms and
  cluster analysis}.{\BBCQ}
\newblock
\APACjournalVolNumPages{Bulletin of the Seismological Society of
  America}{95}{3}{904--915}.
\PrintBackRefs{\CurrentBib}

\bibitem [\protect \citeauthoryear {%
P\BPBI M.~Shearer%
}{%
P\BPBI M.~Shearer%
}{%
{\protect \APACyear {1997}}%
}]{%
shearer1997improving}
\APACinsertmetastar {%
shearer1997improving}%
\begin{APACrefauthors}%
Shearer, P\BPBI M.%
\end{APACrefauthors}%
\unskip\
\newblock
\APACrefYearMonthDay{1997}{}{}.
\newblock
{\BBOQ}\APACrefatitle {Improving local earthquake locations using the L1 norm
  and waveform cross correlation: Application to the Whittier Narrows,
  California, aftershock sequence} {Improving local earthquake locations using
  the l1 norm and waveform cross correlation: Application to the whittier
  narrows, california, aftershock sequence}.{\BBCQ}
\newblock
\APACjournalVolNumPages{Journal of Geophysical Research: Solid
  Earth}{102}{B4}{8269--8283}.
\PrintBackRefs{\CurrentBib}

\bibitem [\protect \citeauthoryear {%
Shelly%
}{%
Shelly%
}{%
{\protect \APACyear {2020}}%
{\protect \APACexlab {{\protect \BCnt {1}}}}}]{%
shelly2020high}
\APACinsertmetastar {%
shelly2020high}%
\begin{APACrefauthors}%
Shelly, D\BPBI R.%
\end{APACrefauthors}%
\unskip\
\newblock
\APACrefYearMonthDay{2020{\protect \BCnt {1}}}{}{}.
\newblock
{\BBOQ}\APACrefatitle {A high-resolution seismic catalog for the initial 2019
  Ridgecrest earthquake sequence: Foreshocks, aftershocks, and faulting
  complexity} {A high-resolution seismic catalog for the initial 2019
  ridgecrest earthquake sequence: Foreshocks, aftershocks, and faulting
  complexity}.{\BBCQ}
\newblock
\APACjournalVolNumPages{Seismological Research Letters}{91}{4}{1971--1978}.
\PrintBackRefs{\CurrentBib}

\bibitem [\protect \citeauthoryear {%
Shelly%
}{%
Shelly%
}{%
{\protect \APACyear {2020}}%
{\protect \APACexlab {{\protect \BCnt {2}}}}}]{%
shelly2020higha}
\APACinsertmetastar {%
shelly2020higha}%
\begin{APACrefauthors}%
Shelly, D\BPBI R.%
\end{APACrefauthors}%
\unskip\
\newblock
\APACrefYearMonthDay{2020{\protect \BCnt {2}}}{{\APACmonth{01}}}{}.
\newblock
{\BBOQ}\APACrefatitle {A {{High}}-{{Resolution Seismic Catalog}} for the
  {{Initial}} 2019 {{Ridgecrest Earthquake Sequence}}: {{Foreshocks}},
  {{Aftershocks}}, and {{Faulting Complexity}}} {A {{High}}-{{Resolution
  Seismic Catalog}} for the {{Initial}} 2019 {{Ridgecrest Earthquake
  Sequence}}: {{Foreshocks}}, {{Aftershocks}}, and {{Faulting
  Complexity}}}.{\BBCQ}
\newblock
\APACjournalVolNumPages{Seismological Research Letters}{91}{4}{1971--1978}.
\PrintBackRefs{\CurrentBib}

\bibitem [\protect \citeauthoryear {%
Shelly%
, Ellsworth%
\BCBL {}\ \BBA {} Hill%
}{%
Shelly%
\ \protect \BOthers {.}}{%
{\protect \APACyear {2016}}%
}]{%
shelly2016fluid}
\APACinsertmetastar {%
shelly2016fluid}%
\begin{APACrefauthors}%
Shelly, D\BPBI R.%
, Ellsworth, W\BPBI L.%
\BCBL {}\ \BBA {} Hill, D\BPBI P.%
\end{APACrefauthors}%
\unskip\
\newblock
\APACrefYearMonthDay{2016}{}{}.
\newblock
{\BBOQ}\APACrefatitle {Fluid-faulting evolution in high definition: Connecting
  fault structure and frequency-magnitude variations during the 2014 Long
  Valley Caldera, California, earthquake swarm} {Fluid-faulting evolution in
  high definition: Connecting fault structure and frequency-magnitude
  variations during the 2014 long valley caldera, california, earthquake
  swarm}.{\BBCQ}
\newblock
\APACjournalVolNumPages{Journal of Geophysical Research: Solid
  Earth}{121}{3}{1776--1795}.
\PrintBackRefs{\CurrentBib}

\bibitem [\protect \citeauthoryear {%
Skoumal%
, Kaven%
\BCBL {}\ \BBA {} Walter%
}{%
Skoumal%
\ \protect \BOthers {.}}{%
{\protect \APACyear {2019}}%
}]{%
skoumal2019characterizing}
\APACinsertmetastar {%
skoumal2019characterizing}%
\begin{APACrefauthors}%
Skoumal, R\BPBI J.%
, Kaven, J\BPBI O.%
\BCBL {}\ \BBA {} Walter, J\BPBI I.%
\end{APACrefauthors}%
\unskip\
\newblock
\APACrefYearMonthDay{2019}{}{}.
\newblock
{\BBOQ}\APACrefatitle {Characterizing Seismogenic Fault Structures in Oklahoma
  Using a Relocated Template-Matched Catalog} {Characterizing seismogenic fault
  structures in oklahoma using a relocated template-matched catalog}.{\BBCQ}
\newblock
\APACjournalVolNumPages{Seismological Research Letters}{90}{4}{1535--1543}.
\PrintBackRefs{\CurrentBib}

\bibitem [\protect \citeauthoryear {%
Smith%
, Azizzadenesheli%
\BCBL {}\ \BBA {} Ross%
}{%
Smith%
\ \protect \BOthers {.}}{%
{\protect \APACyear {2020}}%
}]{%
smith2020eikonet}
\APACinsertmetastar {%
smith2020eikonet}%
\begin{APACrefauthors}%
Smith, J\BPBI D.%
, Azizzadenesheli, K.%
\BCBL {}\ \BBA {} Ross, Z\BPBI E.%
\end{APACrefauthors}%
\unskip\
\newblock
\APACrefYearMonthDay{2020}{}{}.
\newblock
{\BBOQ}\APACrefatitle {Eikonet: Solving the eikonal equation with deep neural
  networks} {Eikonet: Solving the eikonal equation with deep neural
  networks}.{\BBCQ}
\newblock
\APACjournalVolNumPages{IEEE Transactions on Geoscience and Remote
  Sensing}{59}{12}{10685--10696}.
\PrintBackRefs{\CurrentBib}

\bibitem [\protect \citeauthoryear {%
Smith%
, Ross%
, Azizzadenesheli%
\BCBL {}\ \BBA {} Muir%
}{%
Smith%
\ \protect \BOthers {.}}{%
{\protect \APACyear {2022}}%
}]{%
smith2022hyposvi}
\APACinsertmetastar {%
smith2022hyposvi}%
\begin{APACrefauthors}%
Smith, J\BPBI D.%
, Ross, Z\BPBI E.%
, Azizzadenesheli, K.%
\BCBL {}\ \BBA {} Muir, J\BPBI B.%
\end{APACrefauthors}%
\unskip\
\newblock
\APACrefYearMonthDay{2022}{{\APACmonth{01}}}{}.
\newblock
{\BBOQ}\APACrefatitle {{{HypoSVI}}: {{Hypocentre}} Inversion with {{Stein}}
  Variational Inference and Physics Informed Neural Networks} {{{HypoSVI}}:
  {{Hypocentre}} inversion with {{Stein}} variational inference and physics
  informed neural networks}.{\BBCQ}
\newblock
\APACjournalVolNumPages{Geophysical Journal International}{228}{1}{698--710}.
\PrintBackRefs{\CurrentBib}

\bibitem [\protect \citeauthoryear {%
Sun%
, Ross%
, Zhu%
\BCBL {}\ \BBA {} Azizzadenesheli%
}{%
Sun%
\ \protect \BOthers {.}}{%
{\protect \APACyear {2023}}%
}]{%
sun2023phase}
\APACinsertmetastar {%
sun2023phase}%
\begin{APACrefauthors}%
Sun, H.%
, Ross, Z\BPBI E.%
, Zhu, W.%
\BCBL {}\ \BBA {} Azizzadenesheli, K.%
\end{APACrefauthors}%
\unskip\
\newblock
\APACrefYearMonthDay{2023}{}{}.
\newblock
{\BBOQ}\APACrefatitle {Phase neural operator for multi-station picking of
  seismic arrivals} {Phase neural operator for multi-station picking of seismic
  arrivals}.{\BBCQ}
\newblock
\APACjournalVolNumPages{Geophysical Research Letters}{50}{24}{e2023GL106434}.
\PrintBackRefs{\CurrentBib}

\bibitem [\protect \citeauthoryear {%
Szeliski%
}{%
Szeliski%
}{%
{\protect \APACyear {2022}}%
}]{%
szeliski2022computer}
\APACinsertmetastar {%
szeliski2022computer}%
\begin{APACrefauthors}%
Szeliski, R.%
\end{APACrefauthors}%
\unskip\
\newblock
\APACrefYear{2022}.
\newblock
\APACrefbtitle {Computer vision: algorithms and applications} {Computer vision:
  algorithms and applications}.
\newblock
\APACaddressPublisher{}{Springer Nature}.
\PrintBackRefs{\CurrentBib}

\bibitem [\protect \citeauthoryear {%
Thurber%
}{%
Thurber%
}{%
{\protect \APACyear {1983}}%
}]{%
thurber1983earthquake}
\APACinsertmetastar {%
thurber1983earthquake}%
\begin{APACrefauthors}%
Thurber, C\BPBI H.%
\end{APACrefauthors}%
\unskip\
\newblock
\APACrefYearMonthDay{1983}{}{}.
\newblock
{\BBOQ}\APACrefatitle {Earthquake Locations and Three-Dimensional Crustal
  Structure in the Coyote Lake Area, Central California} {Earthquake locations
  and three-dimensional crustal structure in the coyote lake area, central
  california}.{\BBCQ}
\newblock
\APACjournalVolNumPages{Journal of Geophysical Research: Solid
  Earth}{88}{B10}{8226--8236}.
\PrintBackRefs{\CurrentBib}

\bibitem [\protect \citeauthoryear {%
Thurber%
}{%
Thurber%
}{%
{\protect \APACyear {1985}}%
}]{%
thurber1985nonlinear}
\APACinsertmetastar {%
thurber1985nonlinear}%
\begin{APACrefauthors}%
Thurber, C\BPBI H.%
\end{APACrefauthors}%
\unskip\
\newblock
\APACrefYearMonthDay{1985}{{\APACmonth{06}}}{}.
\newblock
{\BBOQ}\APACrefatitle {Nonlinear Earthquake Location: Theory and Examples}
  {Nonlinear earthquake location: Theory and examples}.{\BBCQ}
\newblock
\APACjournalVolNumPages{Bulletin of the Seismological Society of
  America}{75}{3}{779--790}.
\PrintBackRefs{\CurrentBib}

\bibitem [\protect \citeauthoryear {%
Thurber%
}{%
Thurber%
}{%
{\protect \APACyear {1992}}%
}]{%
thurber1992hypocentervelocity}
\APACinsertmetastar {%
thurber1992hypocentervelocity}%
\begin{APACrefauthors}%
Thurber, C\BPBI H.%
\end{APACrefauthors}%
\unskip\
\newblock
\APACrefYearMonthDay{1992}{{\APACmonth{12}}}{}.
\newblock
{\BBOQ}\APACrefatitle {Hypocenter-Velocity Structure Coupling in Local
  Earthquake Tomography} {Hypocenter-velocity structure coupling in local
  earthquake tomography}.{\BBCQ}
\newblock
\APACjournalVolNumPages{Physics of the Earth and Planetary
  Interiors}{75}{1}{55--62}.
\PrintBackRefs{\CurrentBib}

\bibitem [\protect \citeauthoryear {%
Tong%
}{%
Tong%
}{%
{\protect \APACyear {2021}}%
}]{%
tong2021adjoint}
\APACinsertmetastar {%
tong2021adjoint}%
\begin{APACrefauthors}%
Tong, P.%
\end{APACrefauthors}%
\unskip\
\newblock
\APACrefYearMonthDay{2021}{}{}.
\newblock
{\BBOQ}\APACrefatitle {Adjoint-state traveltime tomography: Eikonal
  equation-based methods and application to the Anza area in southern
  California} {Adjoint-state traveltime tomography: Eikonal equation-based
  methods and application to the anza area in southern california}.{\BBCQ}
\newblock
\APACjournalVolNumPages{Journal of Geophysical Research: Solid
  Earth}{126}{5}{e2021JB021818}.
\PrintBackRefs{\CurrentBib}

\bibitem [\protect \citeauthoryear {%
Torr%
\ \BBA {} Zisserman%
}{%
Torr%
\ \BBA {} Zisserman%
}{%
{\protect \APACyear {2000}}%
}]{%
torr2000mlesac}
\APACinsertmetastar {%
torr2000mlesac}%
\begin{APACrefauthors}%
Torr, P\BPBI H.%
\BCBT {}\ \BBA {} Zisserman, A.%
\end{APACrefauthors}%
\unskip\
\newblock
\APACrefYearMonthDay{2000}{}{}.
\newblock
{\BBOQ}\APACrefatitle {MLESAC: A new robust estimator with application to
  estimating image geometry} {Mlesac: A new robust estimator with application
  to estimating image geometry}.{\BBCQ}
\newblock
\APACjournalVolNumPages{Computer vision and image
  understanding}{78}{1}{138--156}.
\PrintBackRefs{\CurrentBib}

\bibitem [\protect \citeauthoryear {%
Trugman%
}{%
Trugman%
}{%
{\protect \APACyear {2020}}%
}]{%
trugman2020stress}
\APACinsertmetastar {%
trugman2020stress}%
\begin{APACrefauthors}%
Trugman, D\BPBI T.%
\end{APACrefauthors}%
\unskip\
\newblock
\APACrefYearMonthDay{2020}{}{}.
\newblock
{\BBOQ}\APACrefatitle {Stress-drop and source scaling of the 2019 Ridgecrest,
  California, earthquake sequence} {Stress-drop and source scaling of the 2019
  ridgecrest, california, earthquake sequence}.{\BBCQ}
\newblock
\APACjournalVolNumPages{Bulletin of the Seismological Society of
  America}{110}{4}{1859--1871}.
\PrintBackRefs{\CurrentBib}

\bibitem [\protect \citeauthoryear {%
Trugman%
}{%
Trugman%
}{%
{\protect \APACyear {2024}}%
}]{%
trugman2024high}
\APACinsertmetastar {%
trugman2024high}%
\begin{APACrefauthors}%
Trugman, D\BPBI T.%
\end{APACrefauthors}%
\unskip\
\newblock
\APACrefYearMonthDay{2024}{}{}.
\newblock
{\BBOQ}\APACrefatitle {A High-Precision Earthquake Catalog for Nevada} {A
  high-precision earthquake catalog for nevada}.{\BBCQ}
\newblock
\APACjournalVolNumPages{Seismological Research Letters}{95}{6}{3737--3745}.
\PrintBackRefs{\CurrentBib}

\bibitem [\protect \citeauthoryear {%
Trugman%
\ \protect \BOthers {.}}{%
Trugman%
\ \protect \BOthers {.}}{%
{\protect \APACyear {2020}}%
}]{%
trugman2020spatiotemporal}
\APACinsertmetastar {%
trugman2020spatiotemporal}%
\begin{APACrefauthors}%
Trugman, D\BPBI T.%
, McBrearty, I\BPBI W.%
, Bolton, D\BPBI C.%
, Guyer, R\BPBI A.%
, Marone, C.%
\BCBL {}\ \BBA {} Johnson, P\BPBI A.%
\end{APACrefauthors}%
\unskip\
\newblock
\APACrefYearMonthDay{2020}{}{}.
\newblock
{\BBOQ}\APACrefatitle {The Spatiotemporal Evolution of Granular Microslip
  Precursors to Laboratory Earthquakes} {The spatiotemporal evolution of
  granular microslip precursors to laboratory earthquakes}.{\BBCQ}
\newblock
\APACjournalVolNumPages{Geophysical Research Letters}{47}{16}{e2020GL088404}.
\PrintBackRefs{\CurrentBib}

\bibitem [\protect \citeauthoryear {%
Trugman%
\ \BBA {} Shearer%
}{%
Trugman%
\ \BBA {} Shearer%
}{%
{\protect \APACyear {2017}}%
}]{%
trugman2017growclust}
\APACinsertmetastar {%
trugman2017growclust}%
\begin{APACrefauthors}%
Trugman, D\BPBI T.%
\BCBT {}\ \BBA {} Shearer, P\BPBI M.%
\end{APACrefauthors}%
\unskip\
\newblock
\APACrefYearMonthDay{2017}{{\APACmonth{02}}}{}.
\newblock
{\BBOQ}\APACrefatitle {{{GrowClust}}: {{A Hierarchical Clustering Algorithm}}
  for {{Relative Earthquake Relocation}}, with {{Application}} to the {{Spanish
  Springs}} and {{Sheldon}}, {{Nevada}}, {{Earthquake Sequences}}}
  {{{GrowClust}}: {{A Hierarchical Clustering Algorithm}} for {{Relative
  Earthquake Relocation}}, with {{Application}} to the {{Spanish Springs}} and
  {{Sheldon}}, {{Nevada}}, {{Earthquake Sequences}}}.{\BBCQ}
\newblock
\APACjournalVolNumPages{Seismological Research Letters}{88}{2A}{379--391}.
\PrintBackRefs{\CurrentBib}

\bibitem [\protect \citeauthoryear {%
Um%
\ \BBA {} Thurber%
}{%
Um%
\ \BBA {} Thurber%
}{%
{\protect \APACyear {1987}}%
}]{%
um1987fast}
\APACinsertmetastar {%
um1987fast}%
\begin{APACrefauthors}%
Um, J.%
\BCBT {}\ \BBA {} Thurber, C.%
\end{APACrefauthors}%
\unskip\
\newblock
\APACrefYearMonthDay{1987}{}{}.
\newblock
{\BBOQ}\APACrefatitle {A fast algorithm for two-point seismic ray tracing} {A
  fast algorithm for two-point seismic ray tracing}.{\BBCQ}
\newblock
\APACjournalVolNumPages{Bulletin of the Seismological Society of
  America}{77}{3}{972--986}.
\PrintBackRefs{\CurrentBib}

\bibitem [\protect \citeauthoryear {%
Van~Trier%
\ \BBA {} Symes%
}{%
Van~Trier%
\ \BBA {} Symes%
}{%
{\protect \APACyear {1991}}%
}]{%
van1991upwind}
\APACinsertmetastar {%
van1991upwind}%
\begin{APACrefauthors}%
Van~Trier, J.%
\BCBT {}\ \BBA {} Symes, W\BPBI W.%
\end{APACrefauthors}%
\unskip\
\newblock
\APACrefYearMonthDay{1991}{}{}.
\newblock
{\BBOQ}\APACrefatitle {Upwind finite-difference calculation of traveltimes}
  {Upwind finite-difference calculation of traveltimes}.{\BBCQ}
\newblock
\APACjournalVolNumPages{Geophysics}{56}{6}{812--821}.
\PrintBackRefs{\CurrentBib}

\bibitem [\protect \citeauthoryear {%
Vidale%
}{%
Vidale%
}{%
{\protect \APACyear {1988}}%
}]{%
vidale1988finite}
\APACinsertmetastar {%
vidale1988finite}%
\begin{APACrefauthors}%
Vidale, J.%
\end{APACrefauthors}%
\unskip\
\newblock
\APACrefYearMonthDay{1988}{}{}.
\newblock
{\BBOQ}\APACrefatitle {Finite-difference calculation of travel times}
  {Finite-difference calculation of travel times}.{\BBCQ}
\newblock
\APACjournalVolNumPages{Bulletin of the seismological society of
  America}{78}{6}{2062--2076}.
\PrintBackRefs{\CurrentBib}

\bibitem [\protect \citeauthoryear {%
Waldhauser%
}{%
Waldhauser%
}{%
{\protect \APACyear {2001}}%
}]{%
waldhauser2001hypodd}
\APACinsertmetastar {%
waldhauser2001hypodd}%
\begin{APACrefauthors}%
Waldhauser, F.%
\end{APACrefauthors}%
\unskip\
\newblock
\APACrefYearMonthDay{2001}{}{}.
\newblock
\APACrefbtitle {HypoDD-A program to compute double-difference hypocenter
  locations} {Hypodd-a program to compute double-difference hypocenter
  locations}\ \APACbVolEdTR{}{\BTR{}}.
\PrintBackRefs{\CurrentBib}

\bibitem [\protect \citeauthoryear {%
Waldhauser%
\ \BBA {} Ellsworth%
}{%
Waldhauser%
\ \BBA {} Ellsworth%
}{%
{\protect \APACyear {2000}}%
}]{%
waldhauser2000doubledifference}
\APACinsertmetastar {%
waldhauser2000doubledifference}%
\begin{APACrefauthors}%
Waldhauser, F.%
\BCBT {}\ \BBA {} Ellsworth, W\BPBI L.%
\end{APACrefauthors}%
\unskip\
\newblock
\APACrefYearMonthDay{2000}{{\APACmonth{12}}}{}.
\newblock
{\BBOQ}\APACrefatitle {A {{Double-Difference Earthquake Location Algorithm}}:
  {{Method}} and {{Application}} to the {{Northern Hayward Fault}},
  {{California}}} {A {{Double-Difference Earthquake Location Algorithm}}:
  {{Method}} and {{Application}} to the {{Northern Hayward Fault}},
  {{California}}}.{\BBCQ}
\newblock
\APACjournalVolNumPages{Bulletin of the Seismological Society of
  America}{90}{6}{1353--1368}.
\PrintBackRefs{\CurrentBib}

\bibitem [\protect \citeauthoryear {%
Waldhauser%
\ \BBA {} Ellsworth%
}{%
Waldhauser%
\ \BBA {} Ellsworth%
}{%
{\protect \APACyear {2002}}%
}]{%
waldhauser2002fault}
\APACinsertmetastar {%
waldhauser2002fault}%
\begin{APACrefauthors}%
Waldhauser, F.%
\BCBT {}\ \BBA {} Ellsworth, W\BPBI L.%
\end{APACrefauthors}%
\unskip\
\newblock
\APACrefYearMonthDay{2002}{}{}.
\newblock
{\BBOQ}\APACrefatitle {Fault structure and mechanics of the Hayward fault,
  California, from double-difference earthquake locations} {Fault structure and
  mechanics of the hayward fault, california, from double-difference earthquake
  locations}.{\BBCQ}
\newblock
\APACjournalVolNumPages{Journal of Geophysical Research: Solid
  Earth}{107}{B3}{ESE--3}.
\PrintBackRefs{\CurrentBib}

\bibitem [\protect \citeauthoryear {%
Waldhauser%
\ \BBA {} Schaff%
}{%
Waldhauser%
\ \BBA {} Schaff%
}{%
{\protect \APACyear {2008}}%
}]{%
waldhauser2008large}
\APACinsertmetastar {%
waldhauser2008large}%
\begin{APACrefauthors}%
Waldhauser, F.%
\BCBT {}\ \BBA {} Schaff, D\BPBI P.%
\end{APACrefauthors}%
\unskip\
\newblock
\APACrefYearMonthDay{2008}{}{}.
\newblock
{\BBOQ}\APACrefatitle {Large-scale relocation of two decades of Northern
  California seismicity using cross-correlation and double-difference methods}
  {Large-scale relocation of two decades of northern california seismicity
  using cross-correlation and double-difference methods}.{\BBCQ}
\newblock
\APACjournalVolNumPages{Journal of Geophysical Research: Solid
  Earth}{113}{B8}{}.
\PrintBackRefs{\CurrentBib}

\bibitem [\protect \citeauthoryear {%
White%
\ \protect \BOthers {.}}{%
White%
\ \protect \BOthers {.}}{%
{\protect \APACyear {2021}}%
}]{%
white2021detailed}
\APACinsertmetastar {%
white2021detailed}%
\begin{APACrefauthors}%
White, M\BPBI C.%
, Fang, H.%
, Catchings, R\BPBI D.%
, Goldman, M\BPBI R.%
, Steidl, J\BPBI H.%
\BCBL {}\ \BBA {} Ben-Zion, Y.%
\end{APACrefauthors}%
\unskip\
\newblock
\APACrefYearMonthDay{2021}{}{}.
\newblock
{\BBOQ}\APACrefatitle {Detailed traveltime tomography and seismic catalogue
  around the 2019 M w7. 1 Ridgecrest, California, earthquake using dense
  rapid-response seismic data} {Detailed traveltime tomography and seismic
  catalogue around the 2019 m w7. 1 ridgecrest, california, earthquake using
  dense rapid-response seismic data}.{\BBCQ}
\newblock
\APACjournalVolNumPages{Geophysical Journal International}{227}{1}{204--227}.
\PrintBackRefs{\CurrentBib}

\bibitem [\protect \citeauthoryear {%
White%
, Fang%
, Nakata%
\BCBL {}\ \BBA {} Ben-Zion%
}{%
White%
\ \protect \BOthers {.}}{%
{\protect \APACyear {2020}}%
}]{%
white2020pykonal}
\APACinsertmetastar {%
white2020pykonal}%
\begin{APACrefauthors}%
White, M\BPBI C.%
, Fang, H.%
, Nakata, N.%
\BCBL {}\ \BBA {} Ben-Zion, Y.%
\end{APACrefauthors}%
\unskip\
\newblock
\APACrefYearMonthDay{2020}{}{}.
\newblock
{\BBOQ}\APACrefatitle {PyKonal: a Python package for solving the eikonal
  equation in spherical and Cartesian coordinates using the fast marching
  method} {Pykonal: a python package for solving the eikonal equation in
  spherical and cartesian coordinates using the fast marching method}.{\BBCQ}
\newblock
\APACjournalVolNumPages{Seismological Research Letters}{91}{4}{2378--2389}.
\PrintBackRefs{\CurrentBib}

\bibitem [\protect \citeauthoryear {%
Wilding%
, Zhu%
, Ross%
\BCBL {}\ \BBA {} Jackson%
}{%
Wilding%
\ \protect \BOthers {.}}{%
{\protect \APACyear {2023}}%
}]{%
wilding2023magmatic}
\APACinsertmetastar {%
wilding2023magmatic}%
\begin{APACrefauthors}%
Wilding, J\BPBI D.%
, Zhu, W.%
, Ross, Z\BPBI E.%
\BCBL {}\ \BBA {} Jackson, J\BPBI M.%
\end{APACrefauthors}%
\unskip\
\newblock
\APACrefYearMonthDay{2023}{}{}.
\newblock
{\BBOQ}\APACrefatitle {The magmatic web beneath Hawai ‘i} {The magmatic web
  beneath hawai ‘i}.{\BBCQ}
\newblock
\APACjournalVolNumPages{Science}{379}{6631}{462--468}.
\PrintBackRefs{\CurrentBib}

\bibitem [\protect \citeauthoryear {%
Woollam%
, Rietbrock%
, Leitloff%
\BCBL {}\ \BBA {} Hinz%
}{%
Woollam%
\ \protect \BOthers {.}}{%
{\protect \APACyear {2020}}%
}]{%
woollam2020hex}
\APACinsertmetastar {%
woollam2020hex}%
\begin{APACrefauthors}%
Woollam, J.%
, Rietbrock, A.%
, Leitloff, J.%
\BCBL {}\ \BBA {} Hinz, S.%
\end{APACrefauthors}%
\unskip\
\newblock
\APACrefYearMonthDay{2020}{{\APACmonth{07}}}{}.
\newblock
{\BBOQ}\APACrefatitle {{{HEX}}: {{Hyperbolic Event eXtractor}}, a {{Seismic
  Phase Associator}} for {{Highly Active Seismic Regions}}} {{{HEX}}:
  {{Hyperbolic Event eXtractor}}, a {{Seismic Phase Associator}} for {{Highly
  Active Seismic Regions}}}.{\BBCQ}
\newblock
\APACjournalVolNumPages{Seismological Research Letters}{91}{5}{2769--2778}.
\PrintBackRefs{\CurrentBib}

\bibitem [\protect \citeauthoryear {%
Yu%
, Ellsworth%
\BCBL {}\ \BBA {} Beroza%
}{%
Yu%
\ \protect \BOthers {.}}{%
{\protect \APACyear {2024}}%
}]{%
yu2024accuracy}
\APACinsertmetastar {%
yu2024accuracy}%
\begin{APACrefauthors}%
Yu, Y.%
, Ellsworth, W\BPBI L.%
\BCBL {}\ \BBA {} Beroza, G\BPBI C.%
\end{APACrefauthors}%
\unskip\
\newblock
\APACrefYearMonthDay{2024}{}{}.
\newblock
{\BBOQ}\APACrefatitle {Accuracy and precision of earthquake location programs:
  Insights from a synthetic controlled experiment} {Accuracy and precision of
  earthquake location programs: Insights from a synthetic controlled
  experiment}.{\BBCQ}
\newblock
\APACjournalVolNumPages{Seismological Research Letters (submitted),
  year}{}{}{}.
\PrintBackRefs{\CurrentBib}

\bibitem [\protect \citeauthoryear {%
D.~Zhao%
, Hasegawa%
\BCBL {}\ \BBA {} Horiuchi%
}{%
D.~Zhao%
\ \protect \BOthers {.}}{%
{\protect \APACyear {1992}}%
}]{%
zhao1992tomographic}
\APACinsertmetastar {%
zhao1992tomographic}%
\begin{APACrefauthors}%
Zhao, D.%
, Hasegawa, A.%
\BCBL {}\ \BBA {} Horiuchi, S.%
\end{APACrefauthors}%
\unskip\
\newblock
\APACrefYearMonthDay{1992}{}{}.
\newblock
{\BBOQ}\APACrefatitle {Tomographic imaging of P and S wave velocity structure
  beneath northeastern Japan} {Tomographic imaging of p and s wave velocity
  structure beneath northeastern japan}.{\BBCQ}
\newblock
\APACjournalVolNumPages{Journal of Geophysical Research: Solid
  Earth}{97}{B13}{19909--19928}.
\PrintBackRefs{\CurrentBib}

\bibitem [\protect \citeauthoryear {%
H.~Zhao%
}{%
H.~Zhao%
}{%
{\protect \APACyear {2005}}%
}]{%
zhao2005fast}
\APACinsertmetastar {%
zhao2005fast}%
\begin{APACrefauthors}%
Zhao, H.%
\end{APACrefauthors}%
\unskip\
\newblock
\APACrefYearMonthDay{2005}{}{}.
\newblock
{\BBOQ}\APACrefatitle {A Fast Sweeping Method for Eikonal Equations} {A fast
  sweeping method for eikonal equations}.{\BBCQ}
\newblock
\APACjournalVolNumPages{Mathematics of Computation}{74}{250}{603--627}.
\PrintBackRefs{\CurrentBib}

\bibitem [\protect \citeauthoryear {%
H\BHBI w.~Zhou%
}{%
H\BHBI w.~Zhou%
}{%
{\protect \APACyear {1994}}%
}]{%
zhou1994rapid}
\APACinsertmetastar {%
zhou1994rapid}%
\begin{APACrefauthors}%
Zhou, H\BHBI w.%
\end{APACrefauthors}%
\unskip\
\newblock
\APACrefYearMonthDay{1994}{}{}.
\newblock
{\BBOQ}\APACrefatitle {Rapid three-dimensional hypocentral determination using
  a master station method} {Rapid three-dimensional hypocentral determination
  using a master station method}.{\BBCQ}
\newblock
\APACjournalVolNumPages{Journal of Geophysical Research: Solid
  Earth}{99}{B8}{15439--15455}.
\PrintBackRefs{\CurrentBib}

\bibitem [\protect \citeauthoryear {%
Y.~Zhou%
, Ghosh%
, Fang%
, Yue%
\BCBL {}\ \BBA {} Zhou%
}{%
Y.~Zhou%
\ \protect \BOthers {.}}{%
{\protect \APACyear {2023}}%
}]{%
zhou2023construction}
\APACinsertmetastar {%
zhou2023construction}%
\begin{APACrefauthors}%
Zhou, Y.%
, Ghosh, A.%
, Fang, L.%
, Yue, H.%
\BCBL {}\ \BBA {} Zhou, S.%
\end{APACrefauthors}%
\unskip\
\newblock
\APACrefYearMonthDay{2023}{}{}.
\newblock
{\BBOQ}\APACrefatitle {Construction of Long-term Seismic Catalog with Deep
  Learning and Characterization of Preseismic Fault Behavior in the
  Ridgecrest-Coso Region (2008-2019)} {Construction of long-term seismic
  catalog with deep learning and characterization of preseismic fault behavior
  in the ridgecrest-coso region (2008-2019)}.{\BBCQ}
\newblock
\APACjournalVolNumPages{Authorea Preprints}{}{}{}.
\PrintBackRefs{\CurrentBib}

\bibitem [\protect \citeauthoryear {%
L.~Zhu%
, Chuang%
, McClellan%
, Liu%
\BCBL {}\ \BBA {} Peng%
}{%
L.~Zhu%
\ \protect \BOthers {.}}{%
{\protect \APACyear {2021}}%
}]{%
zhu2021multichannel}
\APACinsertmetastar {%
zhu2021multichannel}%
\begin{APACrefauthors}%
Zhu, L.%
, Chuang, L.%
, McClellan, J\BPBI H.%
, Liu, E.%
\BCBL {}\ \BBA {} Peng, Z.%
\end{APACrefauthors}%
\unskip\
\newblock
\APACrefYearMonthDay{2021}{{\APACmonth{07}}}{}.
\newblock
{\BBOQ}\APACrefatitle {A Multi-Channel Approach for Automatic Microseismic
  Event Association Using {{RANSAC-based}} Arrival Time Event Clustering
  ({{RATEC}})} {A multi-channel approach for automatic microseismic event
  association using {{RANSAC-based}} arrival time event clustering
  ({{RATEC}})}.{\BBCQ}
\newblock
\APACjournalVolNumPages{Earthquake Research Advances}{1}{3}{100008}.
\PrintBackRefs{\CurrentBib}

\bibitem [\protect \citeauthoryear {%
L.~Zhu%
, Liu%
\BCBL {}\ \BBA {} McClellan%
}{%
L.~Zhu%
\ \protect \BOthers {.}}{%
{\protect \APACyear {2016}}%
}]{%
zhu2016automatic}
\APACinsertmetastar {%
zhu2016automatic}%
\begin{APACrefauthors}%
Zhu, L.%
, Liu, E.%
\BCBL {}\ \BBA {} McClellan, J.%
\end{APACrefauthors}%
\unskip\
\newblock
\APACrefYearMonthDay{2016}{}{}.
\newblock
{\BBOQ}\APACrefatitle {An Automatic Arrival Time Picking Method Based on
  {{RANSAC}} Curve Fitting} {An automatic arrival time picking method based on
  {{RANSAC}} curve fitting}.{\BBCQ}
\newblock
\BIn{} \APACrefbtitle {78th {{EAGE}} Conference and Exhibition 2016} {78th
  {{EAGE}} conference and exhibition 2016}\ (\BVOL\ 2016, \BPGS\ 1--5).
\newblock
\APACaddressPublisher{}{European Association of Geoscientists \& Engineers}.
\PrintBackRefs{\CurrentBib}

\bibitem [\protect \citeauthoryear {%
L.~Zhu%
\ \protect \BOthers {.}}{%
L.~Zhu%
\ \protect \BOthers {.}}{%
{\protect \APACyear {2017}}%
}]{%
zhu2017estimation}
\APACinsertmetastar {%
zhu2017estimation}%
\begin{APACrefauthors}%
Zhu, L.%
, Liu, E.%
, McClellan, J.%
, Zhao, Y.%
, Li, W.%
, Li, Z.%
\BCBL {}\ \BBA {} Peng, Z.%
\end{APACrefauthors}%
\unskip\
\newblock
\APACrefYearMonthDay{2017}{}{}.
\newblock
{\BBOQ}\APACrefatitle {Estimation of Passive Microseismic Event Location Using
  Random Sampling-Based Curve Fitting} {Estimation of passive microseismic
  event location using random sampling-based curve fitting}.{\BBCQ}
\newblock
\BIn{} \APACrefbtitle {{{SEG}} International Exposition and Annual Meeting}
  {{{SEG}} international exposition and annual meeting}\ (\BPGS\ SEG--2017).
\newblock
\APACaddressPublisher{}{SEG}.
\PrintBackRefs{\CurrentBib}

\bibitem [\protect \citeauthoryear {%
W.~Zhu%
\ \BBA {} Beroza%
}{%
W.~Zhu%
\ \BBA {} Beroza%
}{%
{\protect \APACyear {2019}}%
}]{%
zhu2019phasenet}
\APACinsertmetastar {%
zhu2019phasenet}%
\begin{APACrefauthors}%
Zhu, W.%
\BCBT {}\ \BBA {} Beroza, G\BPBI C.%
\end{APACrefauthors}%
\unskip\
\newblock
\APACrefYearMonthDay{2019}{}{}.
\newblock
{\BBOQ}\APACrefatitle {PhaseNet: a deep-neural-network-based seismic
  arrival-time picking method} {Phasenet: a deep-neural-network-based seismic
  arrival-time picking method}.{\BBCQ}
\newblock
\APACjournalVolNumPages{Geophysical Journal International}{216}{1}{261--273}.
\PrintBackRefs{\CurrentBib}

\bibitem [\protect \citeauthoryear {%
W.~Zhu%
, McBrearty%
, Mousavi%
, Ellsworth%
\BCBL {}\ \BBA {} Beroza%
}{%
W.~Zhu%
\ \protect \BOthers {.}}{%
{\protect \APACyear {2022}}%
}]{%
zhu2022earthquake}
\APACinsertmetastar {%
zhu2022earthquake}%
\begin{APACrefauthors}%
Zhu, W.%
, McBrearty, I\BPBI W.%
, Mousavi, S\BPBI M.%
, Ellsworth, W\BPBI L.%
\BCBL {}\ \BBA {} Beroza, G\BPBI C.%
\end{APACrefauthors}%
\unskip\
\newblock
\APACrefYearMonthDay{2022}{}{}.
\newblock
{\BBOQ}\APACrefatitle {Earthquake phase association using a Bayesian Gaussian
  mixture model} {Earthquake phase association using a bayesian gaussian
  mixture model}.{\BBCQ}
\newblock
\APACjournalVolNumPages{Journal of Geophysical Research: Solid
  Earth}{127}{5}{e2021JB023249}.
\PrintBackRefs{\CurrentBib}

\bibitem [\protect \citeauthoryear {%
W.~Zhu%
, Xu%
, Darve%
\BCBL {}\ \BBA {} Beroza%
}{%
W.~Zhu%
\ \protect \BOthers {.}}{%
{\protect \APACyear {2021}}%
}]{%
zhu2021general}
\APACinsertmetastar {%
zhu2021general}%
\begin{APACrefauthors}%
Zhu, W.%
, Xu, K.%
, Darve, E.%
\BCBL {}\ \BBA {} Beroza, G\BPBI C.%
\end{APACrefauthors}%
\unskip\
\newblock
\APACrefYearMonthDay{2021}{}{}.
\newblock
{\BBOQ}\APACrefatitle {A general approach to seismic inversion with automatic
  differentiation} {A general approach to seismic inversion with automatic
  differentiation}.{\BBCQ}
\newblock
\APACjournalVolNumPages{Computers \& Geosciences}{151}{}{104751}.
\PrintBackRefs{\CurrentBib}

\end{thebibliography}

\end{document}